\documentclass[nofootinbib,aps,superscriptaddress]{revtex4}
\usepackage{amssymb,amsmath,graphicx,color,bbold}
\usepackage{empheq}
\usepackage{tikz}
\usepackage{tikz-cd}
\usetikzlibrary{matrix,arrows,decorations.pathmorphing}
\newcommand{\Comment}[1]{{}}
\definecolor{MyDarkBlue}{rgb}{0.15,0.15,0.45}
\usepackage[linktocpage=true]{hyperref}
\hypersetup{
colorlinks=true,
citecolor=MyDarkBlue,
linkcolor=MyDarkBlue,
urlcolor=MyDarkBlue,
pdfauthor={},
pdftitle={},
pdfsubject={}
}

\newcommand{\be}{\begin{equation}}
\newcommand{\ee}{\end{equation}}
\def\ph{\phantom}
\newcommand{\beqn}{\begin{eqnarray}}
\newcommand{\eeqn}{\end{eqnarray}}
\newcommand{\td}{\mathrm{d}}

\newcommand{\nn}{\nonumber}

\newcommand{\gmn}{g_{\mu\nu}}

\newcommand{\hmn}{h_{\mu\nu}}
\newcommand{\hnew}{\tilde{h}}

\newcommand{\D}{D}

\def\ba{\begin{eqnarray}}
\def\ea{\end{eqnarray}}

\begin{document}

\title{Partially Massless Graviton on Beyond Einstein Spacetimes}

\author{Laura Bernard}
\affiliation{CENTRA, Departamento de F\'{\i}sica, Instituto Superior T{\'e}cnico -- IST, Universidade de Lisboa -- UL, Avenida Rovisco Pais 1, 1049 Lisboa, Portugal}

\author{C\'edric Deffayet}
\affiliation{UPMC-CNRS, UMR7095, Institut d'Astrophysique de Paris, GReCO, 98bis boulevard Arago, F-75014 Paris, France}
\affiliation{IHES, Le Bois-Marie, 35 route de Chartres, F-91440 Bures-sur-Yvette, France}

\author{Kurt Hinterbichler}
\affiliation{CERCA, Department of Physics, Case Western Reserve University, 10900 Euclid Ave, Cleveland, OH 44106, USA}

\author{Mikael von Strauss}
\affiliation{UPMC-CNRS, UMR7095, Institut d'Astrophysique de Paris, GReCO, 98bis boulevard Arago, F-75014 Paris, France}
\affiliation{Nordita, KTH Royal Institute of Technology and Stockholm University, Roslagstullsbacken 23, SE-106 91 Stockholm, Sweden}

\begin{abstract}
We show that a partially massless graviton can propagate on a large set of spacetimes which are not Einstein spacetimes.
Starting from a recently constructed theory for a massive graviton that propagates the correct number of degrees of freedom on an arbitrary spacetime, we first give the full explicit form of the scalar constraint responsible for the absence of a sixth degree of freedom.  We then spell out generic conditions for the constraint to be identically satisfied, so that there is a scalar gauge symmetry which makes the graviton partially massless.  These simplify if one assumes that spacetime is Ricci symmetric. Under this assumption, we find explicit non-Einstein spacetimes allowing for the propagation of a partially
massless graviton.

\end{abstract}

\begin{flushright}
\vspace{10pt} \hfill{NORDITA-2017-21} \vspace{20mm}
\end{flushright}

\maketitle

%%%%%%%%%%%%%%%%%%%%%%%%%%%%%%%%%%%%%%%%%%%%%%%%%%%%%%%
\section{Introduction}\label{sec:intro}
%%%%
%%%%%%%%%%%%%%%%%%%%%%%%%%%%%%%%%%%%%%%%%%%%%%%%%%%

A natural physical question to ask is whether the graviton has a mass.  General relativity predicts that the graviton is a self-interacting massless spin-2 particle, but it is possible that small mass corrections to general relativity are present. This question has received renewed interest since the experimental discovery of the acceleration of the universe~\cite{Perlmutter:1998np,Riess:1998cb}, 
the possibility that this acceleration may be explained by a large distance modification of gravity related to the one appearing when the graviton has a mass~\cite{Dvali:2000hr,Deffayet:2000uy,Deffayet:2001pu}, the better understanding of the so-called Vainshtein mechanism of massive gravity~\cite{Deffayet:2001uk,Babichev:2010jd,Babichev:2009jt}, and the recent theoretical discovery of non-linear ghost free theories of massive spin-2~\cite{deRham:2010ik,deRham:2010kj}~(see~\cite{Hinterbichler:2011tt,deRham:2014zqa,Babichev:2013usa} for reviews).

From the point of view of naturalness, an explanation for the observed acceleration using a mechanism driven by a graviton mass would be an improvement over the standard explanation in terms of a cosmological constant.  This is because a small graviton mass is technically natural~\cite{deRham:2012ew,deRham:2013qqa} whereas a small cosmological constant is not.  But such a mechanism would still leave open the question of why various large contributions to the cosmological constant expected from such sources as phase transitions and heavy particle states do not gravitate (see~\cite{Hinterbichler:2017sbd} for a recent summary of the cosmological status of massive gravity and its extensions).

When spacetime is not flat, the division of particles into massless or massive no longer covers all the possibilities.  On de Sitter (dS) space, there exists the mathematical possibility of gravitons which are neither massive nor massless, but instead propagate a number of degrees of freedom greater than that of a massless graviton but less than that of a massive graviton.  These are called ``partially massless" (PM) gravitons, and they enjoy a scalar gauge invariance responsible for removing one of the degrees of freedom of the fully massive graviton~\cite{Deser:1983tm,Deser:1983mm,Higuchi:1986py,Brink:2000ag,Deser:2001pe,Deser:2001us,Deser:2001wx,Deser:2001xr,Zinoviev:2001dt,Skvortsov:2006at,Skvortsov:2009zu}.  

This extra scalar gauge symmetry fixes the mass of the graviton relative to the background curvature, and hence is a candidate symmetry which could fix the cosmological constant relative to the already small graviton mass, thus explaining why the cosmological constant itself is small.   It is primarily for this reason why 
partially massless fields are of interest cosmologically\footnote{Quite apart from cosmology, partially massless fields appear in holographic duals to conformal field theories describing so-called multi-critical points, see e.g.~\cite{Osborn:2016bev,Guerrieri:2016whh,Nakayama:2016dby,Peli:2016gio,Gwak:2016sma,Brust:2016gjy,Fujimori:2016udq,Gliozzi:2016ysv} for some recent work.  These duals are theories with infinite towers of partially massless fields of all spins~\cite{Bekaert:2013zya,Basile:2014wua,Grigoriev:2014kpa,Alkalaev:2014nsa,Joung:2015jza,Brust:2016zns}.} (see the review~\cite{Schmidt-May:2015vnx}).  For this mechanism to be non-trivial, we need to have it realized in a fully non-linear theory whose graviton is partially massless and can be coupled to massive matter.  This has led to many studies of the properties of the linear theory and possible nonlinear extensions~\cite{Zinoviev:2006im,deRham:2012kf,Hassan:2012gz,Hassan:2012rq,Hassan:2013pca,Deser:2013uy,deRham:2013wv,Zinoviev:2014zka,Garcia-Saenz:2014cwa,Hinterbichler:2014xga,Joung:2014aba,Alexandrov:2014oda,Hassan:2015tba,Hinterbichler:2015nua,Cherney:2015jxp,Gwak:2015vfb,Gwak:2015jdo,Garcia-Saenz:2015mqi,Hinterbichler:2016fgl,Bonifacio:2016blz,Apolo:2016ort,Apolo:2016vkn, Gwak:2016sma}.  As a result of these studies, various obstructions and no-go results to an interacting theory have been found, and at this point there is no known four dimensional example of a non-linear ghost-free theory with a finite number of fields in which a partially massless graviton mode propagates fully non-linearly with a gauge symmetry persisting to all orders.

If such a fully non-linear theory exists through some loophole in the various no-go results, then the partially massless graviton would be the fluctuation of some fully non-linear field around a dS background solution.  We would then expect there to be other solutions which are more general than dS, and expanding around these other solutions we would expect to see a PM graviton propagating around these non-dS backgrounds.  Thus, to get more insight into a putative non-linear theory, we can start by asking about what kinds of backgrounds a PM graviton can propagate on.  For example, in the fully massless case, a massless graviton is only known to propagate on an Einstein space~\cite{Aragone:1979bm,Deser:2006sq}.  This is a clue pointing to the fact that the only two-derivative fully non-linear theory of a massless spin 2 is Einstein gravity~\cite{Deser:1987uk,Boulanger:2000rq}, whose vacuum solutions are precisely Einstein spaces. 

Up until now, the only known backgrounds upon which a PM graviton can propagate are Einstein spaces, and obstructions to propagating on more general spacetimes, under certain assumptions about the possible gravitational couplings, have been presented~\cite{Deser:2012qg,Gover:2014vxa}.  Here, we will relax one of these assumptions and show that there are more general non-Einstein spaces upon which a PM graviton can propagate.

We do this by starting with the construction of~\cite{Bernard:2014bfa,Bernard:2015mkk,Bernard:2015uic}, which shows how to couple a fully massive graviton to an arbitrary background metric in such a way that only the correct five degrees of freedom propagate.  We give the full explicit form of the scalar constraint responsible for eliminating a possible sixth degree of freedom.  We then proceed to ask under what conditions on the background and parameters of the theory this constraint becomes identically satisfied, indicating an extra scalar gauge symmetry that emerges and makes the graviton partially massless.  While we won't be able to solve for the most general such conditions, we will be able to find classes of examples which are not Einstein spaces, and thus go beyond the backgrounds previously thought possible.  This gives a hint that there may be some leeway for a fully interacting partially  massless theory of some kind, and raises the question of what the most general background in such a theory will be.

This paper is organized as follows. In the following section we review some technical properties of a massive and partially massless graviton on curved background spacetimes. In the next section we construct explicitly examples of non Einstein spacetimes on which a partially massless graviton can propagate. 

\textbf{Conventions:} We work in four spacetime dimensions and use the mostly plus metric signature~$(-,+,+,+)$.   The curvature conventions are those of~\cite{Carroll:2004st}.  We (anti) symmetrize tensors with unit weight, e.g.,~$S_{(\mu\nu)} = \frac{1}{2}(S_{\mu\nu}+S_{\nu\mu})$.

%%%%%%%%%%%%%%%%%%%%%%%%%%%%%%%%%%%%%%%%%%%%%%%%%%%%%%%
\section{Massive and partially massless gravitons on curved background spacetimes}\label{sec:MGarbBG}
%%%%%%%%%%%%%%%%%%%%%%%%%%%%%%%%%%%%%%%%%%%%%%%%%%%

%%%%%%%%%%%%%%%%%%%%%%%%%%%%%%%%%%%%%%%%%%%%%%%%%%%%%%%
\subsection{Massive graviton on an Einstein background}\label{subsec:PMEinstein}
%%%%%%%%%%%%%%%%%%%%%%%%%%%%%%%%%%%%%%%%%%%%%%%%%%%
%%%%

Before discussing the theory for a massive graviton on a general curved background, as introduced in~\cite{Bernard:2014bfa,Bernard:2015mkk,Bernard:2015uic}, we first review the well known properties of a massive graviton propagating on a generic Einstein spacetime.\footnote{Such a theory was first formulated and studied for a Minkowski background spacetime by Fierz and Pauli~\cite{Fierz:1939ix}. It was then further generalized to a maximally symmetric and later to a generic Einstein spacetime~\cite{Deser:1983mm,Higuchi:1986py, Bengtsson:1994vn, Porrati:2000cp}.  See also \cite{Nepomechie:1983yq,Buchbinder:1999ar,Faci:2012yg,Achour:2013afa}.}  An Einstein spacetime is one whose metric $g_{\mu\nu}$ obeys
\be \label{DEFREIN}
R_{\mu \nu} = \Lambda\, g_{\mu \nu}\,,
\ee
where $R_{\mu \nu}$ is the Ricci tensor and $\Lambda$ is a constant, the cosmological constant. 

A massive graviton propagating on this background has field equations reading (here and henceforth the symbol~$\simeq$~denotes an on-shell equality)
\be
E_{\mu\nu} \simeq 0\,,
\ee
where $E_{\mu \nu}$ is the field equation operator defined by 
\be\label{FPeqsEinstein}
E_{\mu\nu} \equiv
{\mathcal{\D}}_{\mu\nu}^{\ph\mu\ph\nu\rho\sigma}h_{\rho\sigma}
-\Lambda\left(\hmn-\frac1{2}\gmn h\right)
+\frac{m^2}{2}\left(\hmn
-\gmn h\right)\,.
\ee
Here $h_{\mu \nu}$ is a symmetric rank two covariant tensor representing the graviton field, indices are raised and lowered and traced with the background metric, e.g.~$h=g^{\mu\nu}\hmn$, and the linear kinetic operator is given by 
\be\label{gEinstOp_2}
\mathcal{\D}_{\mu\nu}^{\ph\mu\ph\nu\rho\sigma}h_{\rho\sigma}
\equiv -\tfrac1{2}\Big[\delta^\rho_\mu\delta^\sigma_\nu\nabla^2+g^{\rho\sigma}\nabla_\mu\nabla_\nu 
-\delta^\rho_\mu\nabla^\sigma\nabla_\nu-\delta^\rho_\nu\nabla^\sigma\nabla_\mu
-g_{\mu\nu} g^{\rho\sigma}\nabla^2 +g_{\mu\nu}\nabla^\rho\nabla^\sigma\Big]\,h_{\rho\sigma}\,,
\ee\\
where $\nabla$ denotes the covariant derivative associated with the background metric $g_{\mu \nu}$. 
These field equations derive from an action given by 
\be \label{actionEinstein}
S[h]= -{M_{\mathrm{Pl}}^2}\int d^4 x \sqrt{|g|}\left[
h_{\mu\nu}\mathcal{D}^{\mu\nu\rho\sigma}h_{\rho\sigma}
-\Lambda\left(h_{\mu\nu}h^{\mu\nu}-\frac1{2}h^2\right)
+\frac{m^2}{2}\left(h_{\mu\nu}h^{\mu\nu}-h^2\right)\right]\,.
\ee
Note that the kinetic and $\Lambda$ dependent terms appearing on the right hand side of (\ref{FPeqsEinstein}) come from the linearization around an Einstein spacetime of the usual vacuum Einstein equations with a cosmological constant. 

We now review how to count the degrees of freedom in this model in a covariant way~\cite{Porrati:2000cp}, since we will follow the same pattern in the more complicated case of a general background.  One first 
notices that, due to the Bianchi identities identically satisfied by the kinetic operator, 
\be \nabla^\mu\left[{\mathcal{\D}}_{\mu\nu}^{\ph\mu\ph\nu\rho\sigma}h_{\rho\sigma}
-\Lambda\left(\hmn-\frac1{2}\gmn h\right)\right]=0\,,
\ee
one has from the definition (\ref{FPeqsEinstein})
\be
\nabla^\mu E_{\mu\nu}=\frac{m^2}{2}\left(
\nabla^\mu\hmn-g^{\rho\sigma}\nabla_\nu h_{\rho\sigma}\right)\,,
\ee
resulting in the on-shell relation (assuming $m^2\not=0$)
\be \label{CONSEIN1}
\nabla^\mu\hmn-\nabla_\nu h \simeq 0.
\ee
This relation provides four constraint equations (because they contain at most first order derivatives) for $\hmn$, which are referred to as the vector constraints.
These eliminate four degrees of freedom from the original 10 components of the symmetric tensor $\hmn$, leaving six.

To get down to five, the correct count for a massive spin-2 field, we need to find an additional scalar constraint.   
A second covariant divergence of the field equations gives the identity
\be\label{FPdiv}
\nabla^\mu\nabla^\nu E_{\mu\nu}=\frac{m^2}{2}\left(
\nabla^\mu\nabla^\nu\hmn-\nabla^2h\right)\,.
\ee
If, on the other hand, we trace the field operator \eqref{FPeqsEinstein} with the metric we get,
\be\label{FPtrace}
g^{\mu\nu} E_{\mu\nu}=
\nabla^2h-\nabla^\mu\nabla^\nu\hmn
+\left(\Lambda-\frac{3m^2}{2}\right)h\,.
\ee
By comparing~\eqref{FPdiv} and~\eqref{FPtrace} one sees that the linear combination,
\be\label{FPcomb}
2\nabla^\mu\nabla^\nu  E_{\mu\nu}+
m^2g^{\mu\nu} E_{\mu\nu}=
\frac{m^2}{2}\left(2\Lambda-3m^2\right)h\,,
\ee
does not contain any second order derivatives (and in fact no first order derivatives either), and hence 
constitutes a scalar constraint reading (assuming $m^2\not=2\Lambda/3$)
\be \label{SCACONSEIN}
\left(2\Lambda-3m^2\right)h \simeq 0.
\ee
For general parameters this constraint implies $h\simeq 0$ which together with (\ref{CONSEIN1}) simplifies the vector constraint into $\nabla^\mu\hmn\simeq0$ and shows that $\hmn$ is transverse-traceless in vacuum.  By enforcing these constraints the equations of motion~\eqref{FPeqsEinstein} are reduced to the following system,
\be\label{FP_GlF}
\left(\nabla^2-m^2\right)\hmn
+2R_{\mu\ph\rho\nu}^{\ph\mu\rho\ph\nu\sigma}h_{\rho\sigma} \simeq 0\,,\qquad
\nabla^\mu\hmn \simeq 0\,,\qquad h \simeq 0\,.
\ee
Hence, on generic Einstein spacetime the above theory describes a massive graviton with five degrees of freedom. 

An exception to the above arises in two cases: $m^2=0$ and $2\Lambda=3m^2$.  When $m^2=0$, the case of a massless graviton, we do not have any direct constraint \eqref{CONSEIN1} but instead a Noether identity which signals a vector gauge symmetry, which is nothing but the linearized diffeomorphism symmetry of general relativity.   The case $2\Lambda=3m^2$ is when the mass saturates the so-called Higuchi bound~\cite{Higuchi:1986py}. In this case, the linear combination~\eqref{FPcomb} vanishes off-shell, and \eqref{FPcomb} becomes a Noether identity signaling the presence  of a new scalar gauge symmetry~\cite{Deser:2001pe, Deser:2001us}; the field equations are invariant under $h_{\mu \nu} \rightarrow h_{\mu \nu} + \Delta h_{\mu \nu}$, where $\Delta h_{\mu \nu}$ is given by 
\be\label{gaugetrafodS}
\Delta\hmn=\left(\nabla_\mu\nabla_\nu+\frac{m^2}{2}\gmn\right)\xi(x)=\left(\nabla_\mu\nabla_\nu+\frac{\Lambda}{3}\gmn\right)\xi(x)
\,,
\ee
for an arbitrary scalar gauge function $\xi(x)$. This is a symmetry of the quadratic action (\ref{actionEinstein}); for an arbitrary variation $\delta h_{\mu \nu}$ of $h_{\mu \nu}$, the variation of the action  
(\ref{actionEinstein}) is given by
\be
\delta S = -2M_{\mathrm{Pl}}^2 \int d^4 x \sqrt{|g|}\, E^{\mu \nu} \delta h_{\mu \nu}.
\ee
Inserting \eqref{gaugetrafodS} for the variation of $h_{\mu \nu}$,  $\delta h_{\mu \nu} = \Delta h_{\mu \nu}$, integrating by parts, and using the off-shell vanishing of the right hand side of (\ref{FPcomb}) whenever the Higuchi bound is saturated (i.e.~$2\Lambda=3m^2$), we see that $\delta S$ vanishes for the variation $\Delta h_{\mu \nu}$.  A gauge symmetry removes two degrees of freedom (in contrast to a constraint which removes only one), and so we are left with a total of $10-4-2=4$ degrees of freedom. This can be interpreted as a scalar mode becoming a non-propagating pure gauge mode at the enhanced symmetry point, leaving just the four helicity modes ($\pm2, \pm1$).  This is a partially massless graviton.

%%%%%%%%%%%%%%%%%%%%%%%%%%%%%%%%%%%%%%%%%%%%%%%%%%%%%%%
\subsection{Massive graviton on an arbitrary background}\label{subsec:action}
%%%%%%%%%%%%%%%%%%%%%%%%%%%%%%%%%%%%%%%%%%%%%%%%%%%
%%%%

Going beyond Einstein backgrounds, a theory  for a massive graviton $h_{\mu\nu}$ on an arbitrary background spacetime with metric $g_{\mu \nu}$ has been obtained in Refs.~\cite{Bernard:2014bfa,Bernard:2015mkk,Bernard:2015uic}. 
It is written using a tensor $S_{\mu \nu}$ which is defined in terms of the Ricci curvature $R_{\mu \nu}$ of the metric by the relation\footnote{If we define another rank two tensor, $f_{\mu \nu}$, out of the metric $g_{\mu \nu}$ and the tensor $S_{\mu \nu}$ by the following relation
\ba
f_{\mu \nu} = g_{\mu \sigma} S^{\sigma}_{\hphantom{\sigma} \rho} S^{\rho}_{\hphantom{\rho}\nu},
\ea
then this tensor is what is usually called the "reference metric" in the framework of the de Rham, Gabadadze and Tolley (dRGT) theory~\cite{deRham:2010kj,deRham:2010ik,deRham:2011rn,Hassan:2011ea,Hassan:2011zd,Hassan:2011tf,Hassan:2011hr}. However, we stress that we use here a point of view where this reference metric plays no role since we only consider a linear theory defined on a spacetime endowed with a single metric $g_{\mu \nu}$ and all quantities of interest concerning the background spacetimes (e.g. the curvature tensor, the covariant derivatives...) are defined with respect to $g_{\mu\nu}$.} 
\begin{equation} \label{relRSgeneral}
 R^\mu_{~\nu} = m^2\left[ \left(\beta_0 + \frac{1}{2} e_1 \beta_1 \right) \delta^\mu_\nu + \left(\beta_1 + \beta_2 e_1\right) S^\mu_{~\nu} - \beta_2 (S^2)^\mu_{~\nu}\right]\,.
\end{equation}
where $\beta_0$, $\beta_1$ and $\beta_2$ are dimensionless parameters, and $\left(S^2\right)^\mu_{~\nu}$ denotes the square of the tensor $S^\mu_{~\nu}$ considered as a matrix, $\left(S^2\right)^\mu_{~\nu} = S^\mu_{~\rho} S^{\rho}_{~\nu}.$  
Note that the tensor $S_{\mu \nu}$ is symmetric. This can be seen in several ways, one being to notice that \eqref{relRSgeneral} can be formally solved for $S_{\mu\nu}$ and obtain an infinite series consisting of powers of $R_{\mu\nu}$.  Since $R_{\mu\nu}$ is symmetric and powers of a symmetric matrix are symmetric, $S_{\mu\nu}$ is also symmetric, as well as algebraic in $R_{\mu\nu}$. 
 Also, since the inverse of a symmetric matrix is symmetric, $(S^{-1})^{\mu \nu}$, when it exists, is also symmetric. The theory is expressed using the elementary symmetric polynomials $e_n(S)$, $n=0,\cdots,4$
\ba\label{e1e2def}
e_0&=& 1 \, ,\\
e_1&=&S^\rho_{~\rho} \, ,\\
 e_2&=&\frac1{2}\left(S^\rho_{~\rho}S^\nu_{~\nu}
-S^\rho_{~\nu}S^\nu_{~\rho}\right)  \, ,\\
e_3 &=& \frac1{6}\left(S^\rho_{~\rho}S^\nu_{~\nu}S^\mu_{~\mu}
-3S^\mu_{~\mu}S^\rho_{~\nu}S^\nu_{~\rho} +2S^\rho_{~\nu}S^\nu_{~\mu}S^\mu_{~\rho}\right)  \, ,\\
e_4 &=& \det(S)  \, .
\ea

The theory obtained in~\cite{Bernard:2014bfa,Bernard:2015mkk,Bernard:2015uic} for a massive graviton $\hmn$ on an arbitrary background has the vacuum field equations 
\begin{align}\label{lineqs} E_{\mu \nu} \equiv 
\mathcal{E}_{\mu\nu}^{\ph{\mu\nu}\rho\sigma}h_{\rho\sigma}
+\frac{m^2}{2}\biggl[&
2\left(\beta_0+\beta_1e_1+\beta_2e_2\right)\hmn
-\left(\beta_1+\beta_2e_1\right)\left(h_{\mu\rho}S^\rho_{~\nu}+h_{\nu\rho}S^\rho_{~\mu}\right)\nn\\
&-\left(\beta_1\gmn+\beta_2e_1\gmn-\beta_2S_{\mu\nu}\right)h_{\rho\sigma}S^{\rho\sigma}
+\beta_2\gmn h_{\rho\sigma}(S^2)^{\rho\sigma}\nn\\
&-\left(\beta_1+\beta_2e_1\right)\left(g_{\mu\rho}\delta S^\rho_{~\nu}
+g_{\nu\rho}\delta S^\rho_{~\mu}\right)\biggr]\simeq 0\,,
\end{align}
where 
we have defined the linearized Einstein operator
\begin{align}\label{gEinstOp_1}
\mathcal{E}_{\mu\nu}^{\ph{\mu\nu}\rho\sigma}h_{\rho\sigma}
\equiv-\frac1{2}\biggl[&\delta^\rho_\mu\delta^\sigma_\nu\nabla^2
+g^{\rho\sigma}\nabla_\mu\nabla_\nu-\delta^\rho_\mu\nabla^\sigma\nabla_\nu
-\delta^\rho_\nu\nabla^\sigma\nabla_\mu-g_{\mu\nu}g^{\rho\sigma}\nabla^2
+g_{\mu\nu}\nabla^\rho\nabla^\sigma\nonumber\\
&+\delta^\rho_\mu\delta^\sigma_\nu R-g_{\mu\nu}R^{\rho\sigma}\biggr]\,h_{\rho\sigma}\,,
\end{align}
and the
 tensor $\delta S$ has components given by 
\begin{align}\label{d_Sm}
\delta S^{\lambda}_{\hphantom{\lambda}\mu}=\,\, &
\dfrac{1}{2}\,g^{\nu \lambda} \Bigl[\, e_{4}\,c_{1}\,\Bigl(
\delta_{\nu}^{\rho}\delta^{\sigma}_{\mu}+\delta_{\nu}^{\sigma}\delta_{\mu}^{\rho}-g_{\mu\nu}g^{\rho\sigma}
\Bigr) + e_{4}\,c_{2}\,\Bigl(
S_{\nu}^{\rho}\delta^{\sigma}_{\mu}+S_{\nu}^{\sigma}\delta_{\mu}^{\rho}-S_{\mu\nu}g^{\rho\sigma}
-\gmn S^{\rho\sigma}
\Bigr) \nn\\
&- e_{3}\,c_{1}\,\Bigl(
\delta_{\nu}^{\rho}S^{\sigma}_{\mu}+\delta_{\nu}^{\sigma}S_{\mu}^{\rho} \Bigr) 
+\left(e_{2}\,c_{1}-e_{4}\,c_{3}+e_{3}\,c_{2}\right)\,S_{\mu\nu}S^{\rho\sigma}
\nn\\[1pt]
& +e_{4}\,c_{3}\,\Bigl[\delta^{\sigma}_{\mu}[S^{2}]_{\nu}^{\rho} 
+\delta_{\mu}^{\rho}[S^{2}]_{\nu}^{\sigma}-g^{\rho\sigma}[S^{2}]_{\mu\nu}+\delta_{\nu}^{\rho}[S^{2}]^{\sigma}_{\mu}+\delta_{\nu}^{\sigma}[S^{2}]_{\mu}^{\rho}-g_{\mu\nu}[S^{2}]^{\rho\sigma}
\Bigr] \nn\\[1pt]
& - e_{3}\,c_{2}\,\Bigl(
S_{\nu}^{\rho}S^{\sigma}_{\mu}+S_{\nu}^{\sigma}S_{\mu}^{\rho} \Bigr)
%\\[1pt]
%& 
-e_{3}\,c_{3}\,\Bigl(
S^{\sigma}_{\mu}[S^{2}]_{\nu}^{\rho}+S_{\mu}^{\rho}[S^{2}]_{\nu}^{\sigma}+S_{\nu}^{\rho}[S^{2}]^{\sigma}_{\mu}+S_{\nu}^{\sigma}[S^{2}]_{\mu}^{\rho}
\Bigr) \nn\\[1pt]
& +\left(e_{3}\,c_{3}-e_{1}\,c_{1}\right)\,\left(
S^{\rho\sigma}[S^{2}]_{\mu\nu}+S_{\mu\nu}[S^{2}]^{\rho\sigma} \right)
 - \left(c_{1}-e_{2}\,c_{3}\right)\,\Bigl(
[S^{2}]_{\nu}^{\rho}[S^{2}]^{\sigma}_{\mu}+[S^{2}]_{\nu}^{\sigma}[S^{2}]_{\mu}^{\rho}
\Bigr) \nn\\[1pt]
&
+c_{4}\,[S^{2}]_{\mu\nu}[S^{2}]^{\rho\sigma}
+c_{1}\,\Bigl([S^{3}]_{\mu\nu}S^{\rho\sigma}+S_{\mu\nu}[S^{3}]^{\rho\sigma}\Bigr)
+c_{2}\Bigl([S^{3}]_{\mu\nu}[S^{2}]^{\rho\sigma}+[S^{2}]_{\mu\nu}[S^{3}]^{\rho\sigma}\Bigr)\nn\\
&+ c_{3}\,[S^{3}]_{\mu\nu}[S^{3}]^{\rho\sigma} \,\Bigr]\,h_{\rho\sigma} \,,
\end{align}
with the coefficients $c_i$ are given by

\begin{align}
c_{1} &= \dfrac{e_{3}-e_{1}e_{2}}{-e_{1}e_{2}e_{3}+e_{3}^{2}+e_{1}^{2}e_{4}} \,,\;
 c_{2} = \dfrac{e_{1}^{2}}{-e_{1}e_{2}e_{3}+e_{3}^{2}+e_{1}^{2}e_{4}} \,, \;\nn\\
c_{3}& = \dfrac{-e_{1}}{-e_{1}e_{2}e_{3}+e_{3}^{2}+e_{1}^{2}e_{4}} \,,\;
c_{4} = \dfrac{e_{3}-e^3_{1}}{-e_{1}e_{2}e_{3}+e_{3}^{2}+e_{1}^{2}e_{4}} \,.\; \label{cnrelationse}
\end{align}
This theory can only be formulated if the tensor $S$ is such that the spectrum of eigenvalues of $S$, $\sigma(S)$, and the spectrum of its negative, $\sigma(-S)$, do not intersect, i.e.~we should have $\sigma(S)\cap\sigma(-S)=\emptyset$. This is generically the case and will be checked for the solutions we present further down. This additionally ensures that the denominators in Eqs.~\eqref{cnrelationse} do not vanish. { It also implies} that the tensor $S^{\lambda}_{\hphantom{\lambda}\mu}$, considered as a matrix, is invertible which will be used later. Since the equations are linear in $h_{\mu\nu}$, the theory has a quadratic action given by 
\be
S =-{M_{\mathrm{Pl}}^{2}}\int\mathrm{d}^{4}x\sqrt{|g|}\,h^{\mu\nu}E_{\mu\nu}. 
\ee

Note that the above theory was derived in refs.~\cite{Bernard:2014bfa,Bernard:2015mkk,Bernard:2015uic} by linearizing the non-linear ghost-free massive gravity theory of de Rham, Gabadadze and Tolley~\cite{deRham:2010kj} (dRGT) on an arbitrary background~\cite{Hassan:2011vm}.  However, the details of this derivation are inessential for our present purposes and will not be used here. Rather, we will just use the linear theory as defined here as our starting point.

%%%%%%%%%%%%%%%%%%%%%%%%%%%%%%%%%%%%%%%%%%%%%%%%%%%%%%%
\subsection{Covariant constraint counting on a general background}\label{subsec:PMconditions1}
%%%%%%%%%%%%%%%%%%%%%%%%%%%%%%%%%%%%%%%%%%%%%%%%%%%
%%%%

A key feature of this theory is that, as shown using purely covariant constraints in~\cite{Bernard:2014bfa,Bernard:2015mkk,Bernard:2015uic}, it  always propagates at most 5 degrees of freedom on generic backgrounds and hence it is free from the linearized version of the ``Boulware-Deser" ghost~\cite{Boulware:1973my} that was present in previous constructions of non linear massive gravity or non Fierz-Pauli linear theories~\cite{Fierz:1939ix}.  This covariant constraint analysis is crucial for us, as it will allow us to formulate generic covariant conditions for partial masslessness.

The analysis parallels the analysis we reviewed in section~\ref{subsec:PMEinstein} for the case of Einstein spacetimes.  One first notes that, because $\mathcal{E}_{\mu\nu}^{\ph{\mu\nu}\rho\sigma}h_{\rho\sigma}$ is the linearization of the Einstein tensor $\mathcal{G}_{\mu\nu}$, the Bianchi identity $\nabla^\mu\mathcal{G}_{\mu\nu}=0$ implies that
\be
\nabla^\mu E_{\mu \nu} \simeq 0\, 
\ee
does not involve second order derivatives of $h_{\mu\nu}$ and thus gives vector constraints which are the generalization of Eq.~\eqref{CONSEIN1}. These generalized vector constraints eliminate four degrees of freedom of the graviton. 

An extra (scalar) constraint can be found~\cite{Bernard:2014bfa, Bernard:2015mkk,Bernard:2015uic} and reads 
\be\label{scalarC}
\mathcal{C}\equiv
(S^{-1})^\nu_{~\rho}\nabla^\rho\nabla^\mu E_{\mu\nu}
+\frac{m^2\beta_1}{2}g^{\mu\nu} E_{\mu\nu}
+m^2\beta_2S^{\mu\nu}  E_{\mu\nu} \simeq 0\,,
\ee
where we emphasize that the equality on the right hand side only holds on-shell, while the fact that this is a constraint relies on the crucial property that the left hand side of this equality has been shown to depend off-shell only on undifferentiated or once differentiated $h_{\mu \nu}$ but not on second or higher order derivatives of $h_{\mu \nu}$.

We now give the full off-shell expression of this constraint. It is most easily expressed using new field variables $\hnew_{\mu \nu}$ defined through the following equation
\begin{equation*}
h_{\mu\nu} = \Bigl( S^{\lambda}_{~\mu}\delta^{\beta}_{\nu}+S^{\lambda}_{~\nu}\delta^{\beta}_{\mu} \Bigr)\hnew_{\beta\lambda} \,.
\end{equation*}
This, for a given $\hmn$, leads to a unique $\hnew_{\mu \nu}$ provided that $S$ obeys the same condition that allows the theory to be well defined, i.e.~that $\sigma(S)\cap\sigma(-S)=\emptyset$ \cite{Bernard:2015uic}.
In terms of the new variable $\hnew_{\mu \nu}$, the scalar constraint reads
\be \label{GENERICCONS}
\mathcal{C} = m^2 \Bigl[\left(A^{\beta\lambda}+\tilde{A}^{\beta\lambda}\right)\hnew_{\beta\lambda} +B^{\beta\lambda}_{\rho}\,\nabla^{\rho}\hnew_{\beta\lambda} \Bigr] \,,
\ee
where,
\begin{align}\label{defA}
A^{\beta\lambda} \equiv m^{2}\,S^{\beta}_{~\rho} \Bigl[& \left(\beta_{0}\beta_{1} +\beta_{0}\beta_{2}e_{1} +\frac{1}{2}\beta_{1}^{2}e_{1} +\frac{1}{2}\beta_{1}\beta_{2}e_{1}^{2}\right)g^{\rho\lambda} +\left(-2\beta_{0}\beta_{2} -\frac{1}{2}\beta_{1}^{2} -\beta_{1}\beta_{2}e_{1} -2\beta_{2}^{2}e_{2}+\beta_{2}^{2}e_{1}^{2}\right)S^{\rho\lambda} \nn\\
& -\left(\beta_{1}\beta_{2}+\beta_{2}^{2}e_{1}\right)[S^{2}]^{\rho\lambda} \Bigr] \,,
\end{align}
\begin{align}\label{defAt} 
\tilde{A}^{\beta\lambda} \equiv &\ {1\over 2} \left(\beta_{1}+\beta_{2}e_{1}\right)[S^{-1}]^{\nu}_{\gamma}\,\Bigl[-\nabla^{\gamma}S^{\rho\lambda}\nabla_{\nu}S^{\beta}_{\rho} + \nabla^{\gamma}S^{\beta}_{\rho}\nabla^{\lambda}S^{\rho}_{\nu} + \nabla^{\gamma}S^{\rho}_{\nu}\nabla^{\lambda}S^{\beta}_{\rho} - \nabla^{\gamma}S_{\rho\nu}\nabla^{\rho}S^{\beta\lambda} \nn\\
& - S^{\rho\lambda}\nabla^{\gamma}\nabla_{\nu}S^{\beta}_{\rho} + S^{\beta}_{\rho}\nabla^{\gamma}\nabla^{\lambda}S^{\rho}_{\nu} \Bigr] + \beta_{2}\,[S^{-1}]^{\nu}_{~\gamma}\Bigl[ S^{\beta}_{\rho}\nabla^{\lambda}S^{\rho}_{\nu}\nabla^{\gamma}e_{1} -S^{\beta}_{\rho}\nabla_{\nu}S^{\rho\lambda}\nabla^{\gamma}e_{1} \nn\\
& +S^{\lambda}_{\rho}\nabla^{\gamma}S^{\beta}_{\mu}\nabla_{\nu}S^{\rho\mu} +S^{\beta}_{\mu}\nabla^{\gamma}S^{\lambda}_{\rho}\nabla_{\nu}S^{\rho\mu}+S^{\lambda}_{\mu}\nabla^{\gamma}S^{\mu\rho}\nabla_{\nu}S^{\beta}_{\rho} +S^{\mu\rho}\nabla^{\gamma}S^{\lambda}_{\mu}\nabla_{\nu}S^{\beta}_{\rho} \nn\\
& -2S^{\beta}_{\mu}\nabla^{\gamma}S^{\mu\lambda}\nabla_{\nu}e_{1} -S^{\rho}_{\mu}\nabla^{\gamma}S^{\mu}_{\nu}\nabla^{\beta}S^{\lambda}_{\rho} -S^{\beta}_{\mu}\nabla^{\gamma}S^{\mu}_{\rho}\nabla^{\lambda}S^{\rho}_{\nu} -S^{\mu}_{\rho}\nabla^{\gamma}S^{\beta}_{\mu}\nabla^{\lambda}S^{\rho}_{\nu} \nn\\
& -S^{\beta}_{\rho}\nabla^{\gamma}S^{\mu}_{\nu}\nabla^{\lambda}S^{\rho}_{\mu} +S^{\beta}_{\mu}\nabla^{\gamma}S^{\mu}_{\nu}\nabla^{\lambda}e_{1} +S^{\mu}_{\rho}\nabla^{\gamma}S_{\mu\nu}\nabla^{\rho}S^{\beta\lambda} -S^{\lambda}_{\mu}\nabla^{\gamma}S^{\mu}_{\nu}\nabla^{\rho}S^{\beta}_{\rho} \nn\\
& -S^{\lambda}_{\mu}\nabla^{\gamma}S^{\beta}_{\rho}\nabla^{\mu}S^{\rho}_{\nu} -S^{\beta}_{\rho}\nabla^{\gamma}S^{\lambda}_{\mu}\nabla^{\mu}S^{\rho}_{\nu} -S^{\lambda}_{\mu}\nabla^{\gamma}S^{\rho}_{\nu}\nabla^{\mu}S^{\beta}_{\rho} +2S^{\beta}_{\mu}\nabla^{\gamma}S^{\mu\lambda}\nabla^{\rho}S_{\rho\nu} \nn\\
&  +2S^{\beta}_{\rho}\nabla^{\gamma}S_{\mu\nu}\nabla^{\mu}S^{\rho\lambda} +S^{\lambda}_{\rho}S^{\beta}_{\mu}\nabla^{\gamma}\nabla_{\nu}S^{\rho\mu} +[S^2]^{\lambda\rho}\nabla^{\gamma}\nabla_{\nu}S^{\beta}_{\rho} -[S^2]^{\beta\lambda}\nabla^{\gamma}\nabla_{\nu}e_{1} \nn\\
& -[S^2]^{\beta}_{\rho}\nabla^{\gamma}\nabla^{\lambda}S^{\rho}_{\nu} -S^{\lambda}_{\mu}S^{\beta}_{\rho}\nabla^{\gamma}\nabla^{\mu}S^{\rho}_{\nu} +[S^2]^{\beta\lambda}\nabla^{\gamma}\nabla^{\rho}S_{\rho\nu} \Bigr] +\beta_{2}\,\Bigl[+\nabla^{\beta}S^{\lambda}_{\gamma}\nabla^{\gamma}e_{1} \nn\\
& -\nabla_{\gamma}S^{\beta\lambda}\nabla^{\gamma}e_{1} -\nabla^{\mu}S^{\rho}_{\mu}\nabla^{\beta}S^{\lambda}_{\rho} -\nabla^{\mu}S^{\beta}_{\rho}\nabla^{\lambda}S^{\rho}_{\mu} +\nabla^{\mu}S^{\beta}_{\mu}\nabla^{\lambda}e_{1} +\nabla_{\mu}S^{\mu}_{\rho}\nabla^{\rho}S^{\beta\lambda} \nn\\
& -\nabla^{\mu}S^{\lambda}_{\mu}\nabla^{\rho}S^{\beta}_{\rho} -\nabla^{\rho}S^{\lambda}_{\mu}\nabla^{\mu}S^{\beta}_{\rho} +2\nabla_{\mu}S^{\beta}_{\rho}\nabla^{\mu}S^{\rho\lambda} -S^{\beta}_{\rho}\nabla^{\lambda}\nabla^{\mu}S^{\rho}_{\mu} +S^{\beta}_{\gamma}\nabla^{\gamma}\nabla^{\lambda}e_{1} \nn\\
& -S^{\lambda}_{\gamma}\nabla^{\gamma}\nabla^{\rho}S^{\beta}_{\rho} +S^{\beta}_{\rho}\nabla^{\gamma}\nabla_{\gamma}S^{\rho\lambda}\Bigr]  + (\beta\leftrightarrow\lambda)\,,
\end{align}
\begin{align}\label{defB}
B^{\beta\lambda}_{\rho} \equiv \ & {1\over 2} \left(\beta_{1}+\beta_{2}e_{1}\right)[S^{-1}]^{\nu}_{\gamma}\,\left[-S^{\sigma\lambda}\delta^{\gamma}_{\rho}\nabla_{\nu}S^{\beta}_{\sigma} + \delta^{\gamma}_{\rho}S^{\beta}_{\sigma}\nabla^{\lambda}S^{\sigma}_{\nu} + \delta^{\lambda}_{\rho}S^{\beta}_{\sigma}\nabla^{\gamma}S^{\sigma}_{\nu} - S^{\beta\lambda}\nabla^{\gamma}S_{\nu\rho} \right] \nn\\
& + \beta_{2}\,[S^{-1}]^{\nu}_{~\gamma}\,\Bigl[ \delta^{\gamma}_{\rho}S^{\lambda}_{\delta}S^{\beta}_{\mu}\nabla_{\nu}S^{\delta\mu} +\delta^{\gamma}_{\rho}[S^2]^{\lambda\mu}\nabla_{\nu}S^{\beta}_{\mu} -\delta^{\gamma}_{\rho}[S^2]^{\beta\lambda}\nabla_{\nu}e_{1} -\delta^{\gamma}_{\rho}[S^2]^{\beta}_{\mu}\nabla^{\lambda}S^{\mu}_{\nu} \nn\\
& -\delta^{\gamma}_{\rho}S^{\lambda}_{\mu}S^{\beta}_{\delta}\nabla^{\mu}S^{\delta}_{\nu} +\delta^{\gamma}_{\rho}[S^2]^{\beta\lambda}\nabla^{\mu}S_{\mu\nu} +S^{\beta\lambda}S^{\mu}_{\rho}\nabla^{\gamma}S_{\mu\nu} +[S^2]^{\beta\lambda}\nabla^{\gamma}S_{\rho\nu} -\delta^{\beta}_{\rho}[S^2]^{\lambda}_{\mu}\nabla^{\gamma}S^{\mu}_{\nu} \nn\\
& -S^{\beta}_{\rho}S^{\lambda}_{\mu}\nabla^{\gamma}S^{\mu}_{\nu} \Bigr] +\beta_{2}\,\Bigl[-S^{\beta}_{\delta}\nabla^{\lambda}S^{\delta}_{\rho} +S^{\beta}_{\rho}\nabla^{\lambda}e_{1} -S^{\lambda}_{\mu}\nabla^{\mu}S^{\beta}_{\rho} +2S^{\beta}_{\delta}\nabla_{\rho}S^{\delta\lambda} +\delta^{\beta}_{\rho}S^{\lambda}_{\gamma}\nabla^{\gamma}e_{1} \nn\\
& -S^{\beta\lambda}\nabla_{\rho}e_{1} +S^{\beta\lambda}\nabla_{\mu}S^{\mu}_{\rho} -\delta^{\beta}_{\rho}S^{\lambda}_{\delta}\nabla^{\mu}S^{\delta}_{\mu} -S^{\beta}_{\rho}\nabla^{\mu}S^{\lambda}_{\mu} \Bigr]  + (\beta\leftrightarrow\lambda)\,.
\end{align}
These tensors are all symmetric in $\beta,\lambda$ .

Given these constraints (vector plus scalar), the theory propagates at most $10-4-1=5$ degrees of freedom on any background spacetime, appropriate for a massive graviton. One can check that the above constraint~\eqref{GENERICCONS} degenerates to the form (\ref{SCACONSEIN}) on Einstein spacetimes.

\subsection{Condition for Partial masslessness on a generic background spacetime}
Following the discussion in section \ref{subsec:PMEinstein} for Einstein spacetimes, we now ask whether the analog of partial masslessness can be found on spacetimes more generic than Einstein. We will not be able to find here the most general spacetimes for which this happens, but we will be able to exhibit for the first time explicit examples of non Einstein background spacetimes allowing partial masslessness.  

This will happen when the expression for the constraint ${\cal C}$ in Eq.~(\ref{GENERICCONS}) vanishes identically off-shell.
The off-shell vanishing of $\mathcal{C}$ automatically implies the existence of a Noether identity and hence a new scalar gauge symmetry. Indeed, in this case the explicit form of ${\cal C}$ as written in Eq.~\eqref{scalarC} shows that 
the action and the equations of motion are invariant under the transformation $h_{\mu \nu} \rightarrow h_{\mu \nu} + \Delta h_{\mu \nu}$, where $\Delta h_{\mu \nu}$ is now given by 
\be\label{gaugetrafo}
\Delta\hmn=\left[(S^{-1})_\mu^{~\rho}\nabla_\rho\nabla_\nu+(S^{-1})_\nu^{~\rho}\nabla_\rho\nabla_\mu
+m^2\beta_1\gmn+2m^2\beta_2S_{\mu\nu}\right]\,\xi(x)\,,
\ee
with $\xi(x)$ an arbitrary scalar gauge function.  The action is quadratic,
\be
S[h]=-{M_{\mathrm{Pl}}^{2}}\int\td^4x\sqrt{|g|}\,h^{\mu\nu} E_{\mu\nu}\,,
\ee
so upon varying and using~\eqref{gaugetrafo}, the invariance of the action follows from the vanishing of ${\cal C}$.   As before, a gauge symmetry removes two degrees of freedom, and so we are left with at most a total of $10-4-2=4$ degrees of freedom.

%%%%%%%%%%%%%%%%%%%%%%%%%%%%%%%%%%%%%%%%%%%%%%%%%%%%%%%
\section{Background spacetimes for partial masslessness}\label{sec: PMRiccisym}
%%%%%%%%%%%%%%%%%%%%%%%%%%%%%%%%%%%%%%%%%%%%%%%%%%%%%%%

We now proceed to find several classes of background spacetimes for which the scalar constraint vanishes identically so that we have partial masslessness.

%%%%%%%%%%%%%%%%%%%%%%%%%%%%%%%%%%%%%%%%%%%%%%%%%%%%%%%%%%%%%%%%%%%
\subsection{General solution for models with vanishing $\beta_2$}
%%%%%%%%%%%%%%%%%%%%%%%%%%%%%%%%%%%%%%%%%%%%%%%%%%%%%%%%%%%%%%%%%%%

Let us first consider the case of models with vanishing $\beta_2$.  In this case, we will find that the only possible backgrounds are Einstein spaces.

The vanishing of $\beta_2$ in turn implies that $\beta_1$ must be non vanishing for a sensible massive graviton theory to be formulated (otherwise, the equation~\eqref{relRSgeneral} does not define a proper tensor $S_{\mu \nu}$). 
For such a theory, ${A}^{\beta \lambda}$, $\tilde{A}^{\beta \lambda}$ and $B_{\rho}^{\beta \lambda}$ are given by 
\begin{subequations}\label{SympAB}
\begin{align}
{A}^{\beta \lambda} &= \beta_1 \frac{m^2}{2} \Bigl[\left(2\beta_0+\beta_1 e_1\right)S^{\beta\lambda} - \beta_1[S^2]^{\beta\lambda} \Bigr]\, , \\
\tilde{A}^{\beta \lambda} &= \frac{1}{2}\beta_1 [S^{-1}]^{\nu}_{\gamma}\left[-\nabla^{\gamma}S^{\rho\lambda}\nabla_{\nu}S^{\beta}_{\rho} + \nabla^{\gamma}S^{\beta}_{\rho}\nabla^{\lambda}S^{\rho}_{\nu} + \nabla^{\gamma}S^{\rho}_{\nu}\nabla^{\lambda}S^{\beta}_{\rho} - \nabla^{\gamma}S_{\rho\nu}\nabla^{\rho}S^{\beta\lambda} - S^{\rho\lambda}\nabla^{\gamma}\nabla_{\nu}S^{\beta}_{\rho} + S^{\beta}_{\rho}\nabla^{\gamma}\nabla^{\lambda}S^{\rho}_{\nu} \right] +\left(\beta\leftrightarrow\lambda\right) ,\ \\ 
B_{\rho}^{\beta \lambda} &= \frac{1}{2}\beta_1 [S^{-1}]^{\nu}_{\gamma}\left[-S^{\sigma\lambda}\delta^{\gamma}_{\rho}\nabla_{\nu}S^{\beta}_{\sigma} + \delta^{\gamma}_{\rho}S^{\beta}_{\sigma}\nabla^{\lambda}S^{\sigma}_{\nu} + \delta^{\lambda}_{\rho}S^{\beta}_{\sigma}\nabla^{\gamma}S^{\sigma}_{\nu} - S^{\beta\lambda}\nabla^{\gamma}S_{\nu\rho} \right] +\left(\beta\leftrightarrow\lambda\right) ,
\end{align}
\end{subequations}
and the relation \eqref{relRSgeneral} reads 
\begin{equation} \label{defSRsymp}
S_{\mu\nu} = \frac{1}{m^2\,\beta_1}\left(R_{\mu\nu}-\frac{1}{6}\,g_{\mu\nu}\,R-\frac{m^2\,\beta_0}{3}\,g_{\mu\nu}\right) \,.
\end{equation}
The identical vanishing of the constraint \eqref{GENERICCONS} implies that we must have 
\begin{equation}
B^{\beta\lambda}_{\rho} = 0\, .
\end{equation}
Now, using the definitions \eqref{SympAB} we find that 
\begin{equation} \label{MSinit}
B^{\beta\lambda}_{\rho} \left(S^{-1}\right)_{\beta \lambda} = \beta_1 [S^{-1}]^{\nu}_{\gamma}\left[ - \delta^{\gamma}_{\rho}\nabla_{\nu}S^{\sigma}_{\sigma} + \delta^{\gamma}_{\rho}\nabla_{\sigma}S^{\sigma}_{\nu} - 3 \nabla^{\gamma}S_{\nu\rho} \right],
\end{equation}
while the relation \eqref{defSRsymp} implies that
\begin{eqnarray}
\nabla_\sigma S^{\sigma}_{\nu} = \frac{1}{m^2 \beta_1}\frac{1}{3} \nabla_\nu R = \nabla_\nu S^{\sigma}_{\sigma}.
\end{eqnarray}
This shows that the first two terms in the bracket of the left hand side of \eqref{MSinit} cancel each other and one is left with 
\begin{equation} \label{MS}
B^{\beta\lambda}_{\rho} \left(S^{-1}\right)_{\beta \lambda} = -3 \beta_1 [S^{-1}]^{\nu}_{\gamma}\left[\nabla^{\gamma}S_{\nu\rho} \right]\, .
\end{equation}
Thus a necessary condition to get partial masslessness is 
\begin{equation} \label{cond1}
 [S^{-1}]^{\nu}_{\gamma}\left[\nabla^{\gamma}S_{\nu\rho} \right] = 0 \, .
\end{equation}
Now, using $
\left[S^{-1}\right]^\nu_\gamma S_{\nu \rho}= g_{\gamma \rho} $,
we get  
\begin{equation}
[S^{-1}]^{\nu}_{\gamma}\left[\nabla^{\gamma}S_{\nu\rho} \right] = - S_{\nu \rho} \left(\nabla^\gamma\left[S^{-1}\right]^\nu_\gamma\right)  = 0\,.
\end{equation}
The last equality means that $\left(\nabla^\gamma\left[S^{-1}\right]^\nu_\gamma\right)$ is a vector in the kernel of the (invertible) matrix $S^\rho_\nu$ and hence it must vanish.  So we must have 
\begin{equation} \label{cond2}
\left(\nabla^\gamma\left[S^{-1}\right]^\nu_\gamma\right) = 0,
\end{equation}
which will be used later.
We first note that \eqref{cond1} implies that the last two terms entering in the definition of $B^{\beta\lambda}_{\rho}$ in Eq.  \eqref{defB} with $\beta_2=0$ vanish, hence we are left with the expression  
\begin{eqnarray}
B^{\beta\lambda}_{\rho} = \frac{1}{2}\beta_1 [S^{-1}]^{\nu}_{\rho}\left[-S^{\sigma\lambda}\nabla_{\nu}S^{\beta}_{\sigma} + S^{\beta}_{\sigma}\nabla^{\lambda}S^{\sigma}_{\nu} \right] +\left(\beta\leftrightarrow\lambda\right) \, ,
\label{Msimp}
\end{eqnarray}
which must vanish when symmetrized over $\beta$ and $\lambda$. Similarly, after using the condition \eqref{cond1} we are left with the expression for $\tilde{A}^{\beta \lambda}$ given by 
\begin{equation} \label{tildeN}
\tilde{A}^{\beta\lambda} = \frac{1}{2}\beta_1 [S^{-1}]^{\nu}_{\gamma}\left[-\nabla^{\gamma}S^{\rho\lambda}\nabla_{\nu}S^{\beta}_{\rho} + \nabla^{\gamma}S^{\beta}_{\rho}\nabla^{\lambda}S^{\rho}_{\nu}  - S^{\rho\lambda}\nabla^{\gamma}\nabla_{\nu}S^{\beta}_{\rho} + S^{\beta}_{\rho}\nabla^{\gamma}\nabla^{\lambda}S^{\rho}_{\nu} \right] +\left(\beta\leftrightarrow\lambda\right) \,.
\end{equation}
Let us then compute 
$\nabla^\rho B^{(\beta\lambda)}_{\rho}$.   The condition \eqref{cond2} implies that the operator $\nabla^\rho$ ``goes through" the prefactor of $S^{-1}$ on the r.h.s. of the above equation \eqref{Msimp}, and we find that when the condition \eqref{cond2} (and the equivalent \eqref{cond1}) is obeyed, one has 
\begin{equation}
\nabla^\rho B^{\beta\lambda}_{\rho} = \frac{1}{2}\beta_1 [S^{-1}]^{\nu}_{\gamma}\left[-\nabla^{\gamma}S^{\rho\lambda}\nabla_{\nu}S^{\beta}_{\rho} + \nabla^{\gamma}S^{\beta}_{\rho}\nabla^{\lambda}S^{\rho}_{\nu}  - S^{\rho\lambda}\nabla^{\gamma}\nabla_{\nu}S^{\beta}_{\rho} + S^{\beta}_{\rho}\nabla^{\gamma}\nabla^{\lambda}S^{\rho}_{\nu} \right] +\left(\beta\leftrightarrow\lambda\right) \,.
\end{equation}
The right hand side of the above is found to be identical to the r.h.s. of \eqref{tildeN}
which means that the vanishing of $B^{\beta\lambda}_{\rho}$ implies the vanishing of $\tilde{A}^{\beta\lambda}$.

After all this, the only remaining condition for partial masslessness is the vanishing of $A^{\beta \lambda}$. Factoring out one power of $S$ and using that $S$ is invertible in the expression for $A^{\beta \lambda}$, this implies that $S_{\mu \nu}$ is proportional to the metric, which in turns shows, using \eqref{defSRsymp}, that the spacetime must be an Einstein spacetime satisfying $R_{
\mu \nu} = \Lambda g_{\mu\nu}$ for some appropriate value of $\Lambda$. We show next that this conclusion does not extend to the more general cases with non-vanishing $\beta_2$.

%%%%%%%%%%%%%%%%%%%%%%%%%%%%%%%%%%%%%%%%%%%%%%%%%%%%%%%
\subsection{Sufficient conditions for partial masslessness in the general case}\label{subsec:PMconditions2}
%%%%%%%%%%%%%%%%%%%%%%%%%%%%%%%%%%%%%%%%%%%%%%%%%%%
%%%%

We now return to the general case with non-vanishing $\beta_2$, and attempt to find classes of partially massless spacetimes beyond Einstein.
Finding the most general spacetimes for which the right hand side of \eqref{GENERICCONS} vanishes identically is a difficult task given in particular the involved form of the derivative expressions in~\eqref{defAt} and~\eqref{defB}. However, one sees that the form of the constraint ${\cal C}$ becomes much simpler if one 
assumes that 
$S_{\mu \nu}$ is covariantly constant, i.e.~that it obeys
\ba \label{CCT}
\nabla_{\rho} S_{\mu \nu} = 0. 
\ea
The background spacetime then admits a covariantly constant tensor, namely $S_{\mu \nu}$.
We will make this assumption in the rest of this section, stressing that this assumption 
severely restricts the spacetime, and is not necessarily a property of the most general solution.  But, as we will see, it allows us to find spacetimes more generic than the Einstein spacetime where the phenomenon of partial masslessness is known to exist. 

There are several important properties that hold whenever $S$ obeys equation \eqref{CCT}.  First, it 
does not imply that $S_{\mu\nu}\propto\gmn$, and second, it obviously implies that all the scalar invariants made out of the tensor $S_{\mu \nu}$ are constant, and hence in particular, we must have that all of $e_1, e_2, e_3$ and $e_4$ are constant. This together with \eqref{CCT} used in the relation \eqref{relRSgeneral} implies that one must also have 
\ba \label{Riccisymmetric}
\nabla_{\rho} R_{\mu \nu} = 0, 
\ea
i.e.~that the Ricci tensor is covariantly constant. This last condition defines what is known as a {\it Ricci symmetric spacetime}. Such spacetimes, which obviously form a subclass of spacetimes with a covariantly constant tensor, have been studied and classified in various contexts (see e.g.~\cite{Hall,Coley,Levy,Eisenhart,Stephani:2003tm}). In the next subsection, we introduce explicitly some properties of the Ricci-symmetric background spacetimes which will be used later to define our partially massless theory.

%%%%%%%%%%%%%%%%%%%%%%%%%%%%%%%%%%%%%%%%%%%%%%%%%%%%%%%
\subsection{Some properties of suitable Ricci symmetric backgrounds}\label{subsec:Ricsymprop}
%%%%%%%%%%%%%%%%%%%%%%%%%%%%%%%%%%%%%%%%%%%%%%%%%%%
%%%%
The geometry of a spacetime admitting a covariantly constant tensor, say $H_{\mu \nu}$, is constrained by the integrability conditions that derive from the vanishing of the covariant derivatives of the covariantly constant tensor, i.e.
\be
\nabla_{\rho} H_{\mu \nu} = 0. \label{dhiszeroe}
\ee
For instance, using that 
\ba
\nabla_{[\mu} \nabla_{\nu]} H_{\rho \lambda}  = R_{\mu \nu \rho}^{\hphantom{\mu \nu \rho} \sigma} H_{\sigma \lambda} + R_{\mu \nu \lambda}^{\hphantom{\mu \nu \lambda} \sigma} H_{\rho \sigma}, 
\ea
and inserting \eqref{dhiszeroe}, 
we find that such a spacetime must obey the following primary integrability conditions
\ba \label{integrabS}
R_{\mu \nu \rho}^{\hphantom{\mu \nu \rho} \sigma} H_{\sigma \lambda} + R_{\mu \nu \lambda}^{\hphantom{\mu \nu \lambda} \sigma} H_{\rho \sigma} = 0.
\ea
For a Ricci symmetric space \eqref{Riccisymmetric}, when the Ricci tensor is covariantly conserved, the above relation becomes 
a non-trivial relation involving only the  curvature tensor,
\ba \label{integrabR}
R_{\mu \nu \rho}^{\hphantom{\mu \nu \rho} \sigma} R_{\sigma \lambda} + R_{\mu \nu \lambda}^{\hphantom{\mu \nu \lambda} \sigma} R_{\rho \sigma} = 0.
\ea

Furthermore, 4-dimensional (simply connected) spacetimes admitting a covariantly constant symmetric tensor field $H_{\mu\nu}$ can be classified as follows. If $H_{\mu\nu}$
 is not a constant multiple of the metric then such spacetimes can be shown to fall in one of the following two categories~\cite{Hall} (see also e.g.~\cite{Coley,Stephani:2003tm})

\begin{enumerate}
\item The spacetime is $2\otimes2$ decomposable, meaning that the metric can be written in the form
\be \label{2by2}
g_{\mu\nu}\td x^\mu\td x^\nu=g_{ab}(x^c)\td x^a\td x^b+g_{ij}(x^k)\td x^i\td x^j\,,
\ee
where here $a,b,c=0,1$ and $i,j,k=2,3$. In this case, a symmetric and idempotent covariantly constant tensor $H_{\mu \nu}$ can be found, i.e.~a symmetric covariantly constant tensor which satisfies in addition   
$H_{\mu\rho}H^{\rho}_{~\nu}=H_{\mu\nu}$ (that such a covariantly constant idempotent and symmetric tensor exists is true also for $1\otimes3$ decomposable spacetimes but then falls under point 2. below). 
For spacetimes (\ref{2by2}), the Ricci tensor is found to obey
\be\label{RicciD}
R^0_{~0}=R^1_{~1}\,,\qquad R^2_{~2}=R^3_{~3}\,,
\ee
with all other components vanishing. This follows from the fact that the metric (\ref{2by2}) is a direct product of two dimensional metrics, and that the Einstein tensor vanishes identically in $d=2$ dimensions so that the $2d$ Ricci tensors are proportional to their metrics.

An explicit example of such a spacetime which is also Ricci symmetric and that will be used later is
\be \label{linetypeD}
g_{\mu\nu}\td x^\mu\td x^\nu=\frac{2\td x^0 \td x^1 }{(1+(R-E)x^0x^1/8)^2}-\frac{2\td x^2 \td x^3 }{(1-(R+E)x^2x^3/8)^2}\,,
\ee
where $(R-E)/4$ is the scalar curvature of the $(x^0, x^1)$-space and $(R+E)/4$ is the scalar curvature of the $(x^2, x^3)$-space, where both $R$ and $E$ are constant.  As the notation suggests, the constant $R$ is just the Ricci scalar, $R^\rho_{~\rho}=R$, as can be seen explicitly from the expression for the Ricci tensor
\be\label{RtypeD}
R^\rho_{~\nu}=\frac{1}{4}\mathrm{diag}\left[(R-E),(R-E),(R+E),(R+E)\right]\,.
\ee
It can be verified that the Ricci tensor satisfies $\nabla_\rho R_{\mu\nu}=0$.
Furthermore one can show, using Eq.~\eqref{RicciD}, that the Ricci tensor obeys the interesting relation
\be
(R^2)^\rho_{~\nu}=\frac{R}{2}R^\rho_{~\nu}-\frac1{16}\left(R^2-E^2\right)\delta^\rho_\nu\,,
\ee
which can also be deduced directly from the relation (\ref{integrabR}).
The Weyl ($W_{\mu\nu\rho\sigma}$) and Bach ($B_{\mu \nu}$) tensors for this spacetime can also easily be computed and are both non vanishing; the explicit form of the Bach tensor is 
\be \label{BachtypeD}
B^\rho_{~\nu}=\frac{ER}{24}\,\mathrm{diag}\left[-1,-1,1,1\right]\,.
\ee
We see that in order for the Bach tensor to be non-zero we need $ER\neq0$.  In contrast, the case $R=0$ and $E\neq 0$ provide examples of Bach flat non Einstein space-times, which will play a role below. Note further that the spacetimes (\ref{linetypeD}) all belong to the Petrov type D spacetimes\footnote{The Petrov classification classifies spacetimes into 6 classes depending on the nature of principal null directions, see e.g.~\cite{Stephani:2003tm}.} and will be referred to in that way in the next section.

\item There exists a covariantly constant vector (CCV), $N^\mu$, in terms of which a covariantly constant tensor can be constructed as $H_{\mu\nu}=N_\mu N_\nu$
(to which one can also add an arbitrary constant times the metric). The CCV can either be spacelike, timelike or null and its existence implies an integrability condition 
similar to~(\ref{integrabS}), given by 
\be\label{integrabN}
R_{\mu \nu \rho}^{\hphantom{\mu \nu \rho} \sigma} N_\sigma = 0.
\ee
This implies that 
\be
R_\mu^{\hphantom{\mu} \nu}N_\nu = 0, 
\ee
which in turn implies that $N_\mu$ is in the kernel of $R_\mu^{\hphantom{\mu} \nu}$. 
Ricci symmetric spacetimes of interest here and belonging to this class have their line element given by 
\be \label{linetypeO}
g_{\mu\nu}\td x^\mu\td x^\nu=\frac{\td x^2+\td y^2-\epsilon\,\td w^2}{\left[1+R\,(x^2+y^2-\epsilon\,w^2)/24\right]^2}+\epsilon\,\td z^2\,,
\ee
where $\epsilon=\pm1$ depending on whether there is a spacelike ($+$) or timelike ($-$) CCV. The Ricci tensor is given by
\be\label{RtypeO}
R^\rho_{~\nu}=\frac{R}{3}\,\mathrm{diag}\left[1,1,1,0\right]\,.
\ee
This satisfies $R^n_{\mu\nu}=(R/3)^nR_{\mu\nu}$. In fact, for these spacetimes we have that
\be\label{RtypeON}
R_{\mu \nu} = \frac{R}{3} \left(g_{\mu \nu}-\epsilon N_\mu N_\nu\right)\,,
\ee
where $N^\mu$ is a CCV of square norm equal to $\epsilon$. In the case where $N^\mu$ is timelike, i.e.~$\epsilon=-1$, the above spacetime is an Einstein static spacetime. 
These spacetime are both Bach flat and conformally flat, i.e.~have vanishing Bach and Weyl tensors. The spacetimes \eqref{linetypeO} all belong to the Petrov type O class and will be referred to in that way in the next section. We note that other Ricci symmetric spacetimes admitting a CCV can be found, in particular for a null CCV certain pp-wave spacetimes of Petrov type N (see e.g.~\cite{Stephani:2003tm}). However, considering some specific examples of this type, we did not find that those allowed for partial masslessness.
\end{enumerate}

To summarize, the Ricci symmetric spacetimes which might admit partial masslessness are either Petrov type D spacetimes (\ref{linetypeD}) or Petrov type O spacetimes (\ref{linetypeO}).  The later type includes the well known Einstein static Universe.
These spacetimes all obey an interesting idempotency-like relation for the Ricci tensor, reading 
\be\label{R2relRicci}
(R^2)^\rho_{~\nu}=r_1\,R^\rho_{~\nu}+r_2\,\delta^\rho_\nu\,,
\ee
where the constant $r_1$ and $r_2$ are given in the different cases by 
\begin{align}\label{r1r2vals}
\mathrm{type\ D:}&\quad r_1=\frac{R}{2}\,,\qquad r_2=-\frac1{16}\left(R^2-E^2\right)\,,\nn\\
\mathrm{type\ O:}&\quad r_1=\frac{R}{3}\,,\qquad r_2=0\,.
\end{align}

\subsection{Ricci symmetric spacetimes admitting partial masslessness}
\subsubsection{Conditions for partial masslessness on Ricci symmetric spacetimes}

In the case where \eqref{CCT} (and hence also \eqref{Riccisymmetric}) is obeyed, and for generic values of $\beta_0$, $\beta_1$ and $\beta_2$, the scalar constraint 
\eqref{GENERICCONS} reads simply $\mathcal{C} = m^2 A^{\beta \lambda} \hnew_{\beta \lambda}$, so that the condition to get a partially massless graviton is the vanishing of $A^{\beta\lambda}$. Assuming that $S$ is invertible and factoring out one power of S we get that this vanishing reads
\begin{equation} \label{generalPMconstraint}
\delta^{\beta}_{\lambda} \biggl( \beta_2 \beta_0 e_1 + \beta_0 \beta_1 + \frac{\beta_1^2}{2} e_1 +\frac{1}{2}\beta_{1}\beta_{2}e_{1}^{2}\biggr)+ S^{\beta}_{\lambda} \biggl(- 2 \beta_2 \beta_0 + \beta_2^2 e_1^2 - 2 \beta_2^2 e_2 - \frac{\beta_1^2}{2} -\beta_{1}\beta_{2}e_{1} \biggr) -(S^2)^{\beta}_{\lambda} \bigl(\beta_{1}\beta_{2}+ e_1 \beta_2^2\bigr) =0\,.
\end{equation}
This equation together with the relation between $S_{\mu \nu} $ and $R_{\mu \nu}$ given by~\eqref{relRSgeneral} form the system of equation we have to solve. 
Note that when $\beta_{1}=0$, taking the trace of Eq.~\eqref{generalPMconstraint} implies that $\beta_0\beta_2e_1=0$, which further implies $\beta_0e_1=0$ as we have seen that $\beta_2=0$ is not admissible. Then as $e_1=0$ or $\beta_0=0$ only yields Einstein spacetime solutions, we now assume that $\beta_{1}\, ,\beta_2\neq 0$. Finally, the last particular case which may gives interesting solutions is $\beta_1+\beta_2 e_1=0$. In the following we will assume that $\beta_1+\beta_2 e_1\neq 0$, but we will include those particular solutions when reviewing all the possible partially-massless spacetimes.

When \eqref{generalPMconstraint} is obeyed, one can in general obtain $(S^2)_{\mu \nu}$ as a linear combination of $S_{\mu \nu}$ and the metric. Using this in turn in equation~\eqref{relRSgeneral}, one obtains a linear relation between the tensor $S_{\mu \nu}$, the Ricci tensor $R_{\mu \nu}$ and the metric $g_{\mu \nu}$, 
\be\label{RtoS}
R^\rho_{~\nu}=\frac{m^2}{\beta_1}\left(-2\beta_0\beta_2+\frac{3}{2}\beta_1^2\right)S^{\rho}_{~\nu} \,.
\ee
This can be used back in equation~\eqref{generalPMconstraint} to obtain a nontrivial relation between the metric tensor, the Ricci tensor and the square of the Ricci tensor. This relation is precisely of the type (\ref{R2relRicci}) found above and explains why Ricci symmetric spacetimes of the kind introduced previously are suitable backgrounds allowing the propagation of a Partially Massless graviton.

We introduce the following notations that will be used later; we define the parameters $u$, $v$ and $c$ by
\be
u\equiv \frac{\beta_{0}\beta_{2}}{\beta_{1}^{2}} \,,\quad v\equiv m^2 \beta_0  \,, \quad \mathrm{and}\quad c\equiv \beta_1^2\left( -4u + 3\right)  \,.
\ee
Note that $u$ and $c$ are dimensionless, while $v$ has mass dimension two. The relation~\eqref{RtoS} between the Ricci tensor and the tensor $S_{\mu \nu}$ can then be written as
\be\label{RSrelUV}
R^\rho_{~\nu}=\frac{m^2c}{2\beta_1}S^{\rho}_{~\nu} \,.
\ee
From this expression it is obvious that we must demand $c\neq0$ in order not to get an Einstein spacetime (i.e.~$R^\rho_{~\nu}\propto\delta^\rho_\nu$). From this linear relation we can also obtain $S^\rho_{~\nu}$ in terms of $R^\rho_{~\nu}$
\be\label{SRrelUV}
S^\rho_{~\nu}=\frac{2\beta_1}{m^2c}R^\rho_{~\nu}\,.
\ee
%and also its square
%\be
%(S^2)^\rho_{~\nu}=\frac{4\beta_0^{2}\beta_1^{2}}{c^2\left(v-2R\right)^{2}}\left[64(R^2)^\rho_{~\nu} -16v R^\rho_{~\nu} +v^2\delta^\rho_{\nu}\right]\,.
%\ee
This allows us to compute any quantity that depends on $S^\rho_{~\nu}$ in terms of $R^\rho_{~\nu}$.
In particular, equation \eqref{generalPMconstraint} can be rewritten as a condition on the curvature, which reads
\be\label{constrR}
\Biggl[v^2\left(4u^{2}-6u+\frac{9}{4}\right) +vR\left(-u+\frac{3}{4}\right)\Biggr] g^{\beta\lambda} +\Biggl[v\left(-4u^{2}+4u-\frac{3}{4}\right) +uR\Biggr]R^{\beta\lambda}  -u[R^{2}]^{\beta\lambda} = 0 \,.
\ee
The two equations \eqref{generalPMconstraint}, \eqref{relRSgeneral} are then equivalent to the pair (\ref{RSrelUV}), (\ref{constrR}) whenever the conditions
\begin{subequations}\label{conditionsbeta}
\beqn
\beta_1\beta_2 &\neq  0,\\
\beta_1+\beta_2 e_1 &\neq 0,\\
v &\neq 0\, ,\\
c &\neq 0\, ,
\eeqn
\end{subequations}
are obeyed. We have checked that whenever one of these conditions is not fulfilled, the only possible solutions are Einstein spacetimes .

A theory admitting Partial Masslessness on a Ricci symmetric background of the kind introduced in subsection~\ref{subsec:Ricsymprop} can thus be searched for, first solving for the $\beta_k$ and $m$ parameters such that~\eqref{constrR} holds (considering that the background spacetime curvature also obeys \eqref{R2relRicci}), and then defining the tensor $S_{\mu \nu}$ from the curvature from the equation \eqref{RSrelUV}.  Specifically, inserting \eqref{R2relRicci} into~\eqref{constrR} we obtain the following two equations to be solved for $u$ and $v$
\be
\left\{\begin{aligned}\label{eqstosolve}
v^2\left(4u^{2}-6u+\frac{9}{4}\right) +vR\left(-u+\frac{3}{4}\right)-ur_{2} &=0 \,,\\
v\left(-4u^{2}+4u-\frac{3}{4}\right) +u\left(R-r_{1}\right) &=0\,.
\end{aligned}\right.
\ee
In order to solve these equations we now restrict to the different Petrov-type spacetimes introduced in section~\ref{subsec:Ricsymprop}.

\subsubsection{Examples of Ricci symmetric spacetimes with partial masslessness}

\paragraph{Petrov type O.} 
In this case, for which we use the line element  \eqref{linetypeO}, we have $r_{1}=\frac{R}{3}$ and $r_{2}=0$, and the Ricci tensor is given by $R^{\mu}_{\nu}=\mathrm{Diag}\left[1,1,1,0\right]$. According to Eq.~\eqref{SRrelUV} the matrix $S$ has to be proportional to the Ricci tensor and it is thus not invertible, which is impossible. Indeed, in this case, the above system \eqref{eqstosolve} does not yield any valid solution with $r_1=R/3$ and $r_2=0$.  As a consequence there is no partially massless solution of type O\footnote{Note, however, that to reach this conclusion we solve the matrix square root
equation \eqref{relRSgeneral} in the trivial way, i.e. by just
taking square roots of the diagonal element of a diagonal matrix.  However it may be that taking a non conventional square root, one which is
not a diagonal matrix, there could be a solution.}.  

%However there exists a class of solutions for the particular case when $\beta_{1}+\beta_{2}e_{1}=0$, corresponding for a given value of the scalar curvature $R$, to the solution such that
%\be\label{soluvO}
%u-\frac{1}{2} > 0\,, \qquad
%m^2\beta_{0} =-\frac{3u}{(2-u)(5-4u)}\left(3-2u\pm\sqrt{u-\frac{1}{2}}\right)R\,,
%\ee
%and
%\be
%S^{\mu}_{\nu}=\pm\frac{\beta_{1}}{\beta_{2}}\sqrt{u-\frac{1}{2}}\delta^{\mu}_{\nu} -\frac{\beta_{1}}{R\beta_{2}}\left(1\pm\sqrt{u-\frac{1}{2}}\right)R^{\mu}_{\nu}\,,
%\ee
%together with the condition $u\neq\frac{3}{4}$ and $u\neq\frac{27}{50}$.

\paragraph{Petrov type D.} In this case we have $r_{1}=\frac{R}{2}$ and $r_{2}=-\frac{1}{16}\left(R^{2}-E^{2}\right)$. The system of equations then reduces to
\be
\left\{\begin{aligned}\label{eqstosolveD}
& v^2\left(4u^{2}-6u+\frac{9}{4}\right) +vR\left(-u\frac{3}{4}\right)-u\frac{E^2-R^2}{16} =0 \,,\\
& v\left(-4u^{2}+4u-\frac{3}{4}\right) +\frac{uR}{2} =0 \,.
\end{aligned}\right.
\ee
We have the solution for $v$,
\be\label{solvD}
v = \frac{2u}{\left(4u-3\right)\left(4u-1\right)} R \,.
\ee
Inserting this into the equations~\eqref{eqstosolveD}, we obtain the following relation between $E$ and $R$
\be\label{solDE2}
E^{2} = \dfrac{\left(4u-3\right)^2}{\left(4u-1\right)^2}R^2\, .
\ee
For given values of $R$ and $E$, $v$ and $u$ are determined by the above two equations~\eqref{solDE2}--\eqref{solvD}, and are given by
\beqn \label{solsol}
u &=& \frac{E -3R}{4(E-R)}, \;\; v = \frac{(E-3R)(E-R)}{8E} \;\; {\rm and} \;\;u =\frac{E +3R}{4(E+R)}, \;\; v = -\frac{(E+3R)(E+R)}{8E}.
\eeqn
Note that there always exist real solutions to these equations.

We also have the particular solutions corresponding to $\beta_{1}+\beta_{2}e_{1}=0$. The conditions are 
\be\label{solDPart}
E^2=2R^2\frac{8u-3}{2(-3+4u)^2}\,, \qquad
v=m^2\beta_{0} =\frac{2u}{-3+4u}R\,,
\ee
and the matrix $S$ is given by
\be
S^{\mu}_{\nu}=-\frac{\beta_{1}}{\beta_{2}}\left(u-\frac{1}{2}\right)\delta^{\mu}_{\nu} +\frac{\beta_{1}}{\beta_{2}R}(-3+4u)R^{\mu}_{\nu}\,.
\ee
The conditions that the matrix $S$ should fulfill, namely that it is invertible and does not have common eigenvalues with $-S$, further implies that,
\be\label{forbidvaluesD}
E\,R\neq 0, \qquad  E\neq \pm R, , \qquad  \frac{E}{R}\neq \frac{2u-1}{4u-3} \qquad  \frac{E}{2R}\neq \pm\frac{u-1}{4u-3} \,.
\ee
These conditions also translate into some forbidden values for $u$ and $v$.
To summarize, for Petrov-type D spacetimes, we have an infinite number of solutions. More precisely, for any given value of the set $(R,E)$ (except for the forbidden values of Eq.~\eqref{forbidvaluesD}), we have up to two different solutions given by Eqs.~\eqref{solDE2} and~\eqref{solvD}, or by Eq.~\eqref{solDPart}.

\section{Conclusions}\label{sec:conclusions}
%%%%%%%%%%%%%%%%%%%%%%%%%%%%%%%%%%%%%%%%%%%%%%%%%%%
%%%%

We have shown that there are non-Einstein backgrounds on which a partially massless graviton can propagate.   We have done this by taking the construction in~\cite{Bernard:2014bfa,Bernard:2015mkk,Bernard:2015uic} of a fully massive graviton propagating on an arbitrary background, and finding backgrounds and values of the parameter for which an additional scalar gauge symmetry emerges.

      Before this, the only known backgrounds on which a partially massless
graviton could propagate were Einstein
spaces. Indeed, it was argued in~\cite{Deser:2012qg} that only Einstein backgrounds can
propagate a PM graviton. There were
certain assumptions going into this argument, and one of these assumptions
was that the mass term for the graviton is at most linear in the curvature
of the background space-time. Our examples here violate
this assumption by having mass terms with arbitrarily high powers of the
curvature of the background metric (if one expands the mass term in
a power series in the curvature).  However, there are
still at most two derivatives acting on the dynamical field as well as on
the background metric. In other words, there are highly non-minimal
curvature couplings
that allow for PM on non Einstein backgrounds.
%Note that this allows not only a PM graviton to propagate on some non
%Einstein backgrounds, but also on such backgrounds (in the case of solutions~\eqref{SolDR0}) with a vanishing Bach tensor,
%which was also argued impossible in~\cite{Deser:2012qg}.

There are several questions that this raises.  One is the nature of the most general backgrounds allowing for a partially massless mode.  We have not attempted to find the most general possible background within the class of models~\cite{Bernard:2014bfa,Bernard:2015mkk,Bernard:2015uic} which propagate a partially massless mode.  We have only found a few restricted classes of backgrounds beyond Einstein spaces, and there may very well be more examples.  It would be interesting to find and characterize the most general possible background, and this would give a clue as to the nature of the backgrounds which may emerge from the equations of motion of a putative non-linear theory. 

Another open question is the nature of propagation of the partially massless graviton about these non-Einstein backgrounds.  The partially massless graviton on dS space is known to propagate exactly luminally~\cite{Deser:1983tm}.  It would be interesting to see whether this continues to be true for these more general backgrounds.   

\vspace{0.5cm}

{\bf Acknowledgment}:   The authors would like to thank Charles Mazuet for pointing out an error in v1 of this manuscript.  KH would like to thank the Institut d'Astrophysique de Paris for hospitality during which this work originated.  The research of CD and MvS leading to these results, as well the visit of KH to the Institut d'Astrophysique de Paris, have received funding from the European Research Council under the European Community's Seventh Framework Programme (FP7/2007-2013 Grant Agreement no. 307934, "NIRG").
L.B. acknowledges financial support provided under the European Union's H2020 ERC Consolidator Grant "Matter and strong-field gravity: New frontiers in Einstein's theory" grant agreement no. MaGRaTh646597. This project has received funding from the European Union's Horizon 2020 research and innovation programme under the Marie Sklodowska-Curie grant agreement No 690904. 
In the process of checking our calculations, we have used the \textit{xTensor} package~\cite{xTensor}
developed by J.-M.~Mart\'{\i}n-Garc\'{\i}a for \textit{Mathematica}.

%%%%%%%%%%%%%%%%%%%%%%%%%%%%%%%%%%%%%%%%%%%%%%%%%%%%%%%%%%%%%%%%%%%
%%%%%%%%%%%%%%%%%%%%%%%%%%%%%%%%%%%%%%%%%%%%%%%%%%%%%%%%%%%%%%%%%%%%%%


\begin{thebibliography}{999}



%\cite{Perlmutter:1998np}
\bibitem{Perlmutter:1998np} 
  S.~Perlmutter {\it et al.} [Supernova Cosmology Project Collaboration],
  %``Measurements of Omega and Lambda from 42 high redshift supernovae,''
  Astrophys.\ J.\  {\bf 517}, 565 (1999)
  doi:10.1086/307221
  [astro-ph/9812133].
  %%CITATION = doi:10.1086/307221;%%
  %9884 citations counted in INSPIRE as of 12 Feb 2017


%\cite{Riess:1998cb}
\bibitem{Riess:1998cb} 
  A.~G.~Riess {\it et al.} [Supernova Search Team],
  %``Observational evidence from supernovae for an accelerating universe and a cosmological constant,''
  Astron.\ J.\  {\bf 116}, 1009 (1998)
  doi:10.1086/300499
  [astro-ph/9805201].
  %%CITATION = doi:10.1086/300499;%%
  %9644 citations counted in INSPIRE as of 12 Feb 2017


%\cite{Dvali:2000hr}
\bibitem{Dvali:2000hr} 
  G.~R.~Dvali, G.~Gabadadze and M.~Porrati,
  %``4-D gravity on a brane in 5-D Minkowski space,''
  Phys.\ Lett.\ B {\bf 485}, 208 (2000)
  doi:10.1016/S0370-2693(00)00669-9
  [hep-th/0005016].
  %%CITATION = doi:10.1016/S0370-2693(00)00669-9;%%
  %2371 citations counted in INSPIRE as of 27 Feb 2017


%\cite{Deffayet:2000uy}
\bibitem{Deffayet:2000uy} 
  C.~Deffayet,
  %``Cosmology on a brane in Minkowski bulk,''
  Phys.\ Lett.\ B {\bf 502}, 199 (2001)
  doi:10.1016/S0370-2693(01)00160-5
  [hep-th/0010186].
  %%CITATION = doi:10.1016/S0370-2693(01)00160-5;%%
  %721 citations counted in INSPIRE as of 27 Feb 2017

%\cite{Deffayet:2001pu}
\bibitem{Deffayet:2001pu} 
  C.~Deffayet, G.~R.~Dvali and G.~Gabadadze,
  %``Accelerated universe from gravity leaking to extra dimensions,''
  Phys.\ Rev.\ D {\bf 65}, 044023 (2002)
  doi:10.1103/PhysRevD.65.044023
  [astro-ph/0105068].
  %%CITATION = doi:10.1103/PhysRevD.65.044023;%%
  %866 citations counted in INSPIRE as of 27 Feb 2017
	
	
	
	
	
	%\cite{Deffayet:2001uk}
\bibitem{Deffayet:2001uk} 
  C.~Deffayet, G.~R.~Dvali, G.~Gabadadze and A.~I.~Vainshtein,
  %``Nonperturbative continuity in graviton mass versus perturbative discontinuity,''
  Phys.\ Rev.\ D {\bf 65}, 044026 (2002)
  doi:10.1103/PhysRevD.65.044026
  [hep-th/0106001].
  %%CITATION = doi:10.1103/PhysRevD.65.044026;%%
  %406 citations counted in INSPIRE as of 27 Feb 2017
	

%\cite{Babichev:2010jd}
\bibitem{Babichev:2010jd}
  E.~Babichev, C.~Deffayet and R.~Ziour,
  %``The Recovery of General Relativity in massive gravity via the Vainshtein mechanism,''
  Phys.\ Rev.\ D {\bf 82} (2010) 104008
  doi:10.1103/PhysRevD.82.104008
  [arXiv:1007.4506 [gr-qc]].
  %%CITATION = doi:10.1103/PhysRevD.82.104008;%%
  %87 citations counted in INSPIRE as of 27 Feb 2017
	
	
%\cite{Babichev:2009jt}
\bibitem{Babichev:2009jt} 
  E.~Babichev, C.~Deffayet and R.~Ziour,
  %``Recovering General Relativity from massive gravity,''
  Phys.\ Rev.\ Lett.\  {\bf 103}, 201102 (2009)
  doi:10.1103/PhysRevLett.103.201102
  [arXiv:0907.4103 [gr-qc]].
  %%CITATION = doi:10.1103/PhysRevLett.103.201102;%%
  %97 citations counted in INSPIRE as of 27 Feb 2017	









%\cite{deRham:2010ik}
\bibitem{deRham:2010ik} 
  C.~de Rham and G.~Gabadadze,
  %``Generalization of the Fierz-Pauli Action,''
  Phys.\ Rev.\ D {\bf 82}, 044020 (2010)
  doi:10.1103/PhysRevD.82.044020
  [arXiv:1007.0443 [hep-th]].
  %%CITATION = doi:10.1103/PhysRevD.82.044020;%%
  %633 citations counted in INSPIRE as of 12 Feb 2017


%\cite{deRham:2010kj}
\bibitem{deRham:2010kj} 
  C.~de Rham, G.~Gabadadze and A.~J.~Tolley,
  %``Resummation of Massive Gravity,''
  Phys.\ Rev.\ Lett.\  {\bf 106}, 231101 (2011)
  doi:10.1103/PhysRevLett.106.231101
  [arXiv:1011.1232 [hep-th]].
  %%CITATION = doi:10.1103/PhysRevLett.106.231101;%%
  %819 citations counted in INSPIRE as of 12 Feb 2017


%\cite{Hinterbichler:2011tt}
\bibitem{Hinterbichler:2011tt} 
  K.~Hinterbichler,
  %``Theoretical Aspects of Massive Gravity,''
  Rev.\ Mod.\ Phys.\  {\bf 84}, 671 (2012)
  doi:10.1103/RevModPhys.84.671
  [arXiv:1105.3735 [hep-th]].
  %%CITATION = doi:10.1103/RevModPhys.84.671;%%
  %475 citations counted in INSPIRE as of 12 Feb 2017


%\cite{deRham:2014zqa}
\bibitem{deRham:2014zqa} 
  C.~de Rham,
  %``Massive Gravity,''
  Living Rev.\ Rel.\  {\bf 17}, 7 (2014)
  doi:10.12942/lrr-2014-7
  [arXiv:1401.4173 [hep-th]].
  %%CITATION = doi:10.12942/lrr-2014-7;%%
  %330 citations counted in INSPIRE as of 12 Feb 2017


%\cite{Babichev:2013usa}
\bibitem{Babichev:2013usa} 
  E.~Babichev and C.~Deffayet,
  %``An introduction to the Vainshtein mechanism,''
  Class.\ Quant.\ Grav.\  {\bf 30}, 184001 (2013)
  doi:10.1088/0264-9381/30/18/184001
  [arXiv:1304.7240 [gr-qc]].
  %%CITATION = doi:10.1088/0264-9381/30/18/184001;%%
  %141 citations counted in INSPIRE as of 27 Feb 2017



%\cite{deRham:2012ew}
\bibitem{deRham:2012ew} 
  C.~de Rham, G.~Gabadadze, L.~Heisenberg and D.~Pirtskhalava,
  %``Nonrenormalization and naturalness in a class of scalar-tensor theories,''
  Phys.\ Rev.\ D {\bf 87}, no. 8, 085017 (2013)
  doi:10.1103/PhysRevD.87.085017
  [arXiv:1212.4128 [hep-th]].
  %%CITATION = doi:10.1103/PhysRevD.87.085017;%%
  %84 citations counted in INSPIRE as of 12 Feb 2017


%\cite{deRham:2013qqa}
\bibitem{deRham:2013qqa} 
  C.~de Rham, L.~Heisenberg and R.~H.~Ribeiro,
  %``Quantum Corrections in Massive Gravity,''
  Phys.\ Rev.\ D {\bf 88}, 084058 (2013)
  doi:10.1103/PhysRevD.88.084058
  [arXiv:1307.7169 [hep-th]].
  %%CITATION = doi:10.1103/PhysRevD.88.084058;%%
  %65 citations counted in INSPIRE as of 12 Feb 2017


%\cite{Hinterbichler:2017sbd}
\bibitem{Hinterbichler:2017sbd} 
  K.~Hinterbichler,
  %``Cosmology of Massive Gravity and its Extensions,''
  Proceedings of the 51st Rencontres de Moriond, ARISF, 2016, ISBN:
  979-10-968-7901-4
  [arXiv:1701.02873 [astro-ph.CO]].
  %%CITATION = ARXIV:1701.02873;%%


%\cite{Deser:1983tm}
\bibitem{Deser:1983tm} 
  S.~Deser and R.~I.~Nepomechie,
  %``Anomalous Propagation of Gauge Fields in Conformally Flat Spaces,''
  Phys.\ Lett.\  {\bf 132B}, 321 (1983).
  doi:10.1016/0370-2693(83)90317-9
  %%CITATION = doi:10.1016/0370-2693(83)90317-9;%%
  %92 citations counted in INSPIRE as of 12 Feb 2017


%\cite{Deser:1983mm}
\bibitem{Deser:1983mm} 
  S.~Deser and R.~I.~Nepomechie,
  %``Gauge Invariance Versus Masslessness in De Sitter Space,''
  Annals Phys.\  {\bf 154}, 396 (1984).
  doi:10.1016/0003-4916(84)90156-8
  %%CITATION = doi:10.1016/0003-4916(84)90156-8;%%
  %213 citations counted in INSPIRE as of 12 Feb 2017


%\cite{Higuchi:1986py}
\bibitem{Higuchi:1986py} 
  A.~Higuchi,
  %``Forbidden Mass Range for Spin-2 Field Theory in De Sitter spacetime,''
  Nucl.\ Phys.\ B {\bf 282}, 397 (1987).
  doi:10.1016/0550-3213(87)90691-2
  %%CITATION = doi:10.1016/0550-3213(87)90691-2;%%
  %294 citations counted in INSPIRE as of 12 Feb 2017


%\cite{Brink:2000ag}
\bibitem{Brink:2000ag} 
  L.~Brink, R.~R.~Metsaev and M.~A.~Vasiliev,
  %``How massless are massless fields in AdS(d),''
  Nucl.\ Phys.\ B {\bf 586}, 183 (2000)
  doi:10.1016/S0550-3213(00)00402-8
  [hep-th/0005136].
  %%CITATION = doi:10.1016/S0550-3213(00)00402-8;%%
  %132 citations counted in INSPIRE as of 12 Feb 2017


%\cite{Deser:2001pe}
\bibitem{Deser:2001pe} 
  S.~Deser and A.~Waldron,
  %``Gauge invariances and phases of massive higher spins in (A)dS,''
  Phys.\ Rev.\ Lett.\  {\bf 87}, 031601 (2001)
  doi:10.1103/PhysRevLett.87.031601
  [hep-th/0102166].
  %%CITATION = doi:10.1103/PhysRevLett.87.031601;%%
  %213 citations counted in INSPIRE as of 12 Feb 2017


%\cite{Deser:2001us}
\bibitem{Deser:2001us} 
  S.~Deser and A.~Waldron,
  %``Partial masslessness of higher spins in (A)dS,''
  Nucl.\ Phys.\ B {\bf 607}, 577 (2001)
  doi:10.1016/S0550-3213(01)00212-7
  [hep-th/0103198].
  %%CITATION = doi:10.1016/S0550-3213(01)00212-7;%%
  %238 citations counted in INSPIRE as of 12 Feb 2017


%\cite{Deser:2001wx}
\bibitem{Deser:2001wx} 
  S.~Deser and A.~Waldron,
  %``Stability of massive cosmological gravitons,''
  Phys.\ Lett.\ B {\bf 508}, 347 (2001)
  doi:10.1016/S0370-2693(01)00523-8
  [hep-th/0103255].
  %%CITATION = doi:10.1016/S0370-2693(01)00523-8;%%
  %146 citations counted in INSPIRE as of 12 Feb 2017


%\cite{Deser:2001xr}
\bibitem{Deser:2001xr} 
  S.~Deser and A.~Waldron,
  %``Null propagation of partially massless higher spins in (A)dS and cosmological constant speculations,''
  Phys.\ Lett.\ B {\bf 513}, 137 (2001)
  doi:10.1016/S0370-2693(01)00756-0
  [hep-th/0105181].
  %%CITATION = doi:10.1016/S0370-2693(01)00756-0;%%
  %136 citations counted in INSPIRE as of 12 Feb 2017


%\cite{Zinoviev:2001dt}
\bibitem{Zinoviev:2001dt} 
  Y.~M.~Zinoviev,
  %``On massive high spin particles in AdS,''
  hep-th/0108192.
  %%CITATION = HEP-TH/0108192;%%
  %139 citations counted in INSPIRE as of 12 Feb 2017


%\cite{Skvortsov:2006at}
\bibitem{Skvortsov:2006at} 
  E.~D.~Skvortsov and M.~A.~Vasiliev,
  %``Geometric formulation for partially massless fields,''
  Nucl.\ Phys.\ B {\bf 756}, 117 (2006)
  doi:10.1016/j.nuclphysb.2006.06.019
  [hep-th/0601095].
  %%CITATION = doi:10.1016/j.nuclphysb.2006.06.019;%%
  %81 citations counted in INSPIRE as of 12 Feb 2017


%\cite{Skvortsov:2009zu}
\bibitem{Skvortsov:2009zu} 
  E.~D.~Skvortsov,
  %``Gauge fields in (A)dS(d) and Connections of its symmetry algebra,''
  J.\ Phys.\ A {\bf 42}, 385401 (2009)
  doi:10.1088/1751-8113/42/38/385401
  [arXiv:0904.2919 [hep-th]].
  %%CITATION = doi:10.1088/1751-8113/42/38/385401;%%
  %40 citations counted in INSPIRE as of 12 Feb 2017


%\cite{Osborn:2016bev}
\bibitem{Osborn:2016bev} 
  H.~Osborn and A.~Stergiou,
  %``C$_{T}$ for non-unitary CFTs in higher dimensions,''
  JHEP {\bf 1606}, 079 (2016)
  doi:10.1007/JHEP06(2016)079
  [arXiv:1603.07307 [hep-th]].
  %%CITATION = doi:10.1007/JHEP06(2016)079;%%
  %7 citations counted in INSPIRE as of 12 Feb 2017


%\cite{Guerrieri:2016whh}
\bibitem{Guerrieri:2016whh} 
  A.~Guerrieri, A.~C.~Petkou and C.~Wen,
  %``The free $\sigma$CFTs,''
  JHEP {\bf 1609}, 019 (2016)
  doi:10.1007/JHEP09(2016)019
  [arXiv:1604.07310 [hep-th]].
  %%CITATION = doi:10.1007/JHEP09(2016)019;%%
  %4 citations counted in INSPIRE as of 12 Feb 2017


%\cite{Nakayama:2016dby}
\bibitem{Nakayama:2016dby} 
  Y.~Nakayama,
  %``Hidden global conformal symmetry without Virasoro extension in theory of elasticity,''
  Annals Phys.\  {\bf 372}, 392 (2016)
  doi:10.1016/j.aop.2016.06.010
  [arXiv:1604.00810 [hep-th]].
  %%CITATION = doi:10.1016/j.aop.2016.06.010;%%
  %5 citations counted in INSPIRE as of 12 Feb 2017


%\cite{Peli:2016gio}
\bibitem{Peli:2016gio} 
  Z.~PŽli, S.~Nagy and K.~Sailer,
  %``Phase structure of the $O(2)$ ghost model with higher-order gradient term,''
  Phys.\ Rev.\ D {\bf 94}, no. 6, 065021 (2016)
  doi:10.1103/PhysRevD.94.065021
  [arXiv:1605.07836 [hep-th]].
  %%CITATION = doi:10.1103/PhysRevD.94.065021;%%
  %4 citations counted in INSPIRE as of 12 Feb 2017


%\cite{Gwak:2016sma}
\bibitem{Gwak:2016sma} 
  S.~Gwak, J.~Kim and S.~J.~Rey,
  %``Massless and Massive Higher Spins from Anti-de Sitter Space Waveguide,''
  JHEP {\bf 1611}, 024 (2016)
  doi:10.1007/JHEP11(2016)024
  [arXiv:1605.06526 [hep-th]].
  %%CITATION = doi:10.1007/JHEP11(2016)024;%%
  %6 citations counted in INSPIRE as of 12 Feb 2017


%\cite{Brust:2016gjy}
\bibitem{Brust:2016gjy} 
  C.~Brust and K.~Hinterbichler,
  %``Free box^k Scalar Conformal Field Theory,''
  arXiv:1607.07439 [hep-th].
  %%CITATION = ARXIV:1607.07439;%%
  %5 citations counted in INSPIRE as of 12 Feb 2017


%\cite{Fujimori:2016udq}
\bibitem{Fujimori:2016udq} 
  T.~Fujimori, M.~Nitta and Y.~Yamada,
  %``Ghostbusters in higher derivative supersymmetric theories: who is afraid of propagating auxiliary fields?,''
  JHEP {\bf 1609}, 106 (2016)
  doi:10.1007/JHEP09(2016)106
  [arXiv:1608.01843 [hep-th]].
  %%CITATION = doi:10.1007/JHEP09(2016)106;%%
  %5 citations counted in INSPIRE as of 12 Feb 2017


%\cite{Gliozzi:2016ysv}
\bibitem{Gliozzi:2016ysv} 
  F.~Gliozzi, A.~Guerrieri, A.~C.~Petkou and C.~Wen,
  %``Generalized Wilson-Fisher critical points from the conformal OPE,''
  arXiv:1611.10344 [hep-th].
  %%CITATION = ARXIV:1611.10344;%%
  %2 citations counted in INSPIRE as of 12 Feb 2017


%\cite{Bekaert:2013zya}
\bibitem{Bekaert:2013zya} 
  X.~Bekaert and M.~Grigoriev,
  %``Higher order singletons, partially massless fields and their boundary values in the ambient approach,''
  Nucl.\ Phys.\ B {\bf 876}, 667 (2013)
  doi:10.1016/j.nuclphysb.2013.08.015
  [arXiv:1305.0162 [hep-th]].
  %%CITATION = doi:10.1016/j.nuclphysb.2013.08.015;%%
  %38 citations counted in INSPIRE as of 12 Feb 2017


%\cite{Basile:2014wua}
\bibitem{Basile:2014wua} 
  T.~Basile, X.~Bekaert and N.~Boulanger,
  %``Flato-Fronsdal theorem for higher-order singletons,''
  JHEP {\bf 1411}, 131 (2014)
  doi:10.1007/JHEP11(2014)131
  [arXiv:1410.7668 [hep-th]].
  %%CITATION = doi:10.1007/JHEP11(2014)131;%%
  %12 citations counted in INSPIRE as of 12 Feb 2017


%\cite{Grigoriev:2014kpa}
\bibitem{Grigoriev:2014kpa} 
  X.~Bekaert and M.~Grigoriev,
  %``Higher-Order Singletons and Partially Massless Fields,''
  Bulg.\ J.\ Phys.\  {\bf 41}, 172 (2014).
  %%CITATION = BJPHD,41,172;%%
  %5 citations counted in INSPIRE as of 12 Feb 2017


%\cite{Alkalaev:2014nsa}
\bibitem{Alkalaev:2014nsa} 
  K.~B.~Alkalaev, M.~Grigoriev and E.~D.~Skvortsov,
  %``Uniformizing higher-spin equations,''
  J.\ Phys.\ A {\bf 48}, no. 1, 015401 (2015)
  doi:10.1088/1751-8113/48/1/015401
  [arXiv:1409.6507 [hep-th]].
  %%CITATION = doi:10.1088/1751-8113/48/1/015401;%%
  %11 citations counted in INSPIRE as of 12 Feb 2017


%\cite{Joung:2015jza}
\bibitem{Joung:2015jza} 
  E.~Joung and K.~Mkrtchyan,
  %``Partially-massless higher-spin algebras and their finite-dimensional truncations,''
  JHEP {\bf 1601}, 003 (2016)
  doi:10.1007/JHEP01(2016)003
  [arXiv:1508.07332 [hep-th]].
  %%CITATION = doi:10.1007/JHEP01(2016)003;%%
  %7 citations counted in INSPIRE as of 12 Feb 2017


%\cite{Brust:2016zns}
\bibitem{Brust:2016zns} 
  C.~Brust and K.~Hinterbichler,
  %``Partially Massless Higher-Spin Theory,''
  arXiv:1610.08510 [hep-th].
  %%CITATION = ARXIV:1610.08510;%%
  %3 citations counted in INSPIRE as of 12 Feb 2017


%\cite{Schmidt-May:2015vnx}
\bibitem{Schmidt-May:2015vnx} 
  A.~Schmidt-May and M.~von Strauss,
  %``Recent developments in bimetric theory,''
  J.\ Phys.\ A {\bf 49}, no. 18, 183001 (2016)
  doi:10.1088/1751-8113/49/18/183001
  [arXiv:1512.00021 [hep-th]].
  %%CITATION = doi:10.1088/1751-8113/49/18/183001;%%
  %33 citations counted in INSPIRE as of 12 Feb 2017


%\cite{Zinoviev:2006im}
\bibitem{Zinoviev:2006im} 
  Y.~M.~Zinoviev,
  %``On massive spin 2 interactions,''
  Nucl.\ Phys.\ B {\bf 770}, 83 (2007)
  doi:10.1016/j.nuclphysb.2007.02.005
  [hep-th/0609170].
  %%CITATION = doi:10.1016/j.nuclphysb.2007.02.005;%%
  %66 citations counted in INSPIRE as of 12 Feb 2017


%\cite{deRham:2012kf}
\bibitem{deRham:2012kf} 
  C.~de Rham and S.~Renaux-Petel,
  %``Massive Gravity on de Sitter and Unique Candidate for Partially Massless Gravity,''
  JCAP {\bf 1301}, 035 (2013)
  doi:10.1088/1475-7516/2013/01/035
  [arXiv:1206.3482 [hep-th]].
  %%CITATION = doi:10.1088/1475-7516/2013/01/035;%%
  %77 citations counted in INSPIRE as of 19 Apr 2017


%\cite{Hassan:2012gz}
\bibitem{Hassan:2012gz} 
  S.~F.~Hassan, A.~Schmidt-May and M.~von Strauss,
  %``On Partially Massless Bimetric Gravity,''
  Phys.\ Lett.\ B {\bf 726}, 834 (2013)
  doi:10.1016/j.physletb.2013.09.021
  [arXiv:1208.1797 [hep-th]].
  %%CITATION = doi:10.1016/j.physletb.2013.09.021;%%
  %56 citations counted in INSPIRE as of 12 Feb 2017


%\cite{Hassan:2012rq}
\bibitem{Hassan:2012rq} 
  S.~F.~Hassan, A.~Schmidt-May and M.~von Strauss,
  %``Bimetric theory and partial masslessness with LanczosÐLovelock terms in arbitrary dimensions,''
  Class.\ Quant.\ Grav.\  {\bf 30}, 184010 (2013)
  doi:10.1088/0264-9381/30/18/184010
  [arXiv:1212.4525 [hep-th]].
  %%CITATION = doi:10.1088/0264-9381/30/18/184010;%%
  %40 citations counted in INSPIRE as of 12 Feb 2017


%\cite{Hassan:2013pca}
\bibitem{Hassan:2013pca} 
  S.~F.~Hassan, A.~Schmidt-May and M.~von Strauss,
  %``Higher Derivative Gravity and Conformal Gravity From Bimetric and Partially Massless Bimetric Theory,''
  Universe {\bf 1}, no. 2, 92 (2015)
  doi:10.3390/universe1020092
  [arXiv:1303.6940 [hep-th]].
  %%CITATION = doi:10.3390/universe1020092;%%
  %48 citations counted in INSPIRE as of 12 Feb 2017


%\cite{Deser:2013uy}
\bibitem{Deser:2013uy} 
  S.~Deser, M.~Sandora and A.~Waldron,
  %``Nonlinear Partially Massless from Massive Gravity?,''
  Phys.\ Rev.\ D {\bf 87}, no. 10, 101501 (2013)
  doi:10.1103/PhysRevD.87.101501
  [arXiv:1301.5621 [hep-th]].
  %%CITATION = doi:10.1103/PhysRevD.87.101501;%%
  %63 citations counted in INSPIRE as of 12 Feb 2017


%\cite{deRham:2013wv}
\bibitem{deRham:2013wv} 
  C.~de Rham, K.~Hinterbichler, R.~A.~Rosen and A.~J.~Tolley,
  %``Evidence for and obstructions to nonlinear partially massless gravity,''
  Phys.\ Rev.\ D {\bf 88}, no. 2, 024003 (2013)
  doi:10.1103/PhysRevD.88.024003
  [arXiv:1302.0025 [hep-th]].
  %%CITATION = doi:10.1103/PhysRevD.88.024003;%%
  %54 citations counted in INSPIRE as of 12 Feb 2017


%\cite{Zinoviev:2014zka}
\bibitem{Zinoviev:2014zka} 
  Y.~M.~Zinoviev,
  %``Massive spin-2 in the FradkinÐVasiliev formalism. I. Partially massless case,''
  Nucl.\ Phys.\ B {\bf 886}, 712 (2014)
  doi:10.1016/j.nuclphysb.2014.07.013
  [arXiv:1405.4065 [hep-th]].
  %%CITATION = doi:10.1016/j.nuclphysb.2014.07.013;%%
  %16 citations counted in INSPIRE as of 12 Feb 2017


%\cite{Garcia-Saenz:2014cwa}
\bibitem{Garcia-Saenz:2014cwa} 
  S.~Garcia-Saenz and R.~A.~Rosen,
  %``A non-linear extension of the spin-2 partially massless symmetry,''
  JHEP {\bf 1505}, 042 (2015)
  doi:10.1007/JHEP05(2015)042
  [arXiv:1410.8734 [hep-th]].
  %%CITATION = doi:10.1007/JHEP05(2015)042;%%
  %17 citations counted in INSPIRE as of 12 Feb 2017


%\cite{Hinterbichler:2014xga}
\bibitem{Hinterbichler:2014xga} 
  K.~Hinterbichler,
  %``Manifest Duality Invariance for the Partially Massless Graviton,''
  Phys.\ Rev.\ D {\bf 91}, no. 2, 026008 (2015)
  doi:10.1103/PhysRevD.91.026008
  [arXiv:1409.3565 [hep-th]].
  %%CITATION = doi:10.1103/PhysRevD.91.026008;%%
  %14 citations counted in INSPIRE as of 12 Feb 2017


%\cite{Joung:2014aba}
\bibitem{Joung:2014aba} 
  E.~Joung, W.~Li and M.~Taronna,
  %``No-Go Theorems for Unitary and Interacting Partially Massless Spin-Two Fields,''
  Phys.\ Rev.\ Lett.\  {\bf 113}, 091101 (2014)
  doi:10.1103/PhysRevLett.113.091101
  [arXiv:1406.2335 [hep-th]].
  %%CITATION = doi:10.1103/PhysRevLett.113.091101;%%
  %36 citations counted in INSPIRE as of 12 Feb 2017


%\cite{Alexandrov:2014oda}
\bibitem{Alexandrov:2014oda} 
  S.~Alexandrov and C.~Deffayet,
  %``On Partially Massless Theory in 3 Dimensions,''
  JCAP {\bf 1503}, no. 03, 043 (2015)
  doi:10.1088/1475-7516/2015/03/043
  [arXiv:1410.2897 [hep-th]].
  %%CITATION = doi:10.1088/1475-7516/2015/03/043;%%
  %7 citations counted in INSPIRE as of 12 Feb 2017


%\cite{Hassan:2015tba}
\bibitem{Hassan:2015tba} 
  S.~F.~Hassan, A.~Schmidt-May and M.~von Strauss,
  %``Extended Weyl Invariance in a Bimetric Model and Partial Masslessness,''
  Class.\ Quant.\ Grav.\  {\bf 33}, no. 1, 015011 (2016)
  doi:10.1088/0264-9381/33/1/015011
  [arXiv:1507.06540 [hep-th]].
  %%CITATION = doi:10.1088/0264-9381/33/1/015011;%%
  %12 citations counted in INSPIRE as of 12 Feb 2017


%\cite{Hinterbichler:2015nua}
\bibitem{Hinterbichler:2015nua} 
  K.~Hinterbichler and R.~A.~Rosen,
  %``Partially Massless Monopoles and Charges,''
  Phys.\ Rev.\ D {\bf 92}, no. 10, 105019 (2015)
  doi:10.1103/PhysRevD.92.105019
  [arXiv:1507.00355 [hep-th]].
  %%CITATION = doi:10.1103/PhysRevD.92.105019;%%
  %7 citations counted in INSPIRE as of 12 Feb 2017


%\cite{Cherney:2015jxp}
\bibitem{Cherney:2015jxp} 
  D.~Cherney, S.~Deser, A.~Waldron and G.~Zahariade,
  %``Non-linear duality invariant partially massless models?,''
  Phys.\ Lett.\ B {\bf 753}, 293 (2016)
  doi:10.1016/j.physletb.2015.12.029
  [arXiv:1511.01053 [hep-th]].
  %%CITATION = doi:10.1016/j.physletb.2015.12.029;%%
  %5 citations counted in INSPIRE as of 12 Feb 2017


%\cite{Gwak:2015vfb}
\bibitem{Gwak:2015vfb} 
  S.~Gwak, E.~Joung, K.~Mkrtchyan and S.~J.~Rey,
  %``Rainbow Valley of Colored (Anti) de Sitter Gravity in Three Dimensions,''
  JHEP {\bf 1604}, 055 (2016)
  doi:10.1007/JHEP04(2016)055
  [arXiv:1511.05220 [hep-th]].
  %%CITATION = doi:10.1007/JHEP04(2016)055;%%
  %10 citations counted in INSPIRE as of 12 Feb 2017


%\cite{Gwak:2015jdo}
\bibitem{Gwak:2015jdo} 
  S.~Gwak, E.~Joung, K.~Mkrtchyan and S.~J.~Rey,
  %``Rainbow vacua of colored higher-spin (A)dS$_{3}$ gravity,''
  JHEP {\bf 1605}, 150 (2016)
  doi:10.1007/JHEP05(2016)150
  [arXiv:1511.05975 [hep-th]].
  %%CITATION = doi:10.1007/JHEP05(2016)150;%%
  %11 citations counted in INSPIRE as of 12 Feb 2017


%\cite{Garcia-Saenz:2015mqi}
\bibitem{Garcia-Saenz:2015mqi} 
  S.~Garcia-Saenz, K.~Hinterbichler, A.~Joyce, E.~Mitsou and R.~A.~Rosen,
  %``No-go for Partially Massless Spin-2 Yang-Mills,''
  JHEP {\bf 1602}, 043 (2016)
  doi:10.1007/JHEP02(2016)043
  [arXiv:1511.03270 [hep-th]].
  %%CITATION = doi:10.1007/JHEP02(2016)043;%%
  %6 citations counted in INSPIRE as of 12 Feb 2017


%\cite{Hinterbichler:2016fgl}
\bibitem{Hinterbichler:2016fgl} 
  K.~Hinterbichler and A.~Joyce,
  %``Manifest Duality for Partially Massless Higher Spins,''
  JHEP {\bf 1609}, 141 (2016)
  doi:10.1007/JHEP09(2016)141
  [arXiv:1608.04385 [hep-th]].
  %%CITATION = doi:10.1007/JHEP09(2016)141;%%
  %4 citations counted in INSPIRE as of 12 Feb 2017


%\cite{Bonifacio:2016blz}
\bibitem{Bonifacio:2016blz} 
  J.~Bonifacio and K.~Hinterbichler,
  %``Kaluza-Klein reduction of massive and partially massless spin-2 fields,''
  Phys.\ Rev.\ D {\bf 95}, no. 2, 024023 (2017)
  doi:10.1103/PhysRevD.95.024023
  [arXiv:1611.00362 [hep-th]].
  %%CITATION = doi:10.1103/PhysRevD.95.024023;%%


%\cite{Apolo:2016ort}
\bibitem{Apolo:2016ort} 
  L.~Apolo and S.~F.~Hassan,
  %``Non-linear partially massless symmetry in an SO(1,5) continuation of conformal gravity,''
  arXiv:1609.09514 [hep-th].
  %%CITATION = ARXIV:1609.09514;%%
  %3 citations counted in INSPIRE as of 12 Feb 2017


%\cite{Apolo:2016vkn}
\bibitem{Apolo:2016vkn} 
  L.~Apolo, S.~F.~Hassan and A.~Lundkvist,
  %``Gauge and global symmetries of the candidate partially massless bimetric gravity,''
  Phys.\ Rev.\ D {\bf 94}, no. 12, 124055 (2016)
  doi:10.1103/PhysRevD.94.124055
  [arXiv:1609.09515 [hep-th]].
  %%CITATION = doi:10.1103/PhysRevD.94.124055;%%
  %5 citations counted in INSPIRE as of 12 Feb 2017


%\cite{Aragone:1979bm}
\bibitem{Aragone:1979bm} 
  C.~Aragone and S.~Deser,
  %``Consistency Problems of Spin-2 Gravity Coupling,''
  Nuovo Cim.\ B {\bf 57}, 33 (1980).
  doi:10.1007/BF02722400
  %%CITATION = doi:10.1007/BF02722400;%%
  %115 citations counted in INSPIRE as of 12 Feb 2017


%\cite{Deser:2006sq}
\bibitem{Deser:2006sq} 
  S.~Deser and M.~Henneaux,
  %``A Note on spin two fields in curved backgrounds,''
  Class.\ Quant.\ Grav.\  {\bf 24}, 1683 (2007)
  doi:10.1088/0264-9381/24/6/N01
  [gr-qc/0611157].
  %%CITATION = doi:10.1088/0264-9381/24/6/N01;%%
  %16 citations counted in INSPIRE as of 12 Feb 2017


%\cite{Deser:1987uk}
\bibitem{Deser:1987uk} 
  S.~Deser,
  %``Gravity From Selfinteraction in a Curved Background,''
  Class.\ Quant.\ Grav.\  {\bf 4}, L99 (1987).
  doi:10.1088/0264-9381/4/4/006
  %%CITATION = doi:10.1088/0264-9381/4/4/006;%%
  %46 citations counted in INSPIRE as of 12 Feb 2017


%\cite{Boulanger:2000rq}
\bibitem{Boulanger:2000rq} 
  N.~Boulanger, T.~Damour, L.~Gualtieri and M.~Henneaux,
  %``Inconsistency of interacting, multigraviton theories,''
  Nucl.\ Phys.\ B {\bf 597}, 127 (2001)
  doi:10.1016/S0550-3213(00)00718-5
  [hep-th/0007220].
  %%CITATION = doi:10.1016/S0550-3213(00)00718-5;%%
  %190 citations counted in INSPIRE as of 12 Feb 2017


%\cite{Deser:2012qg}
\bibitem{Deser:2012qg} 
  S.~Deser, E.~Joung and A.~Waldron,
  %``Partial Masslessness and Conformal Gravity,''
  J.\ Phys.\ A {\bf 46}, 214019 (2013)
  doi:10.1088/1751-8113/46/21/214019
  [arXiv:1208.1307 [hep-th]].
  %%CITATION = doi:10.1088/1751-8113/46/21/214019;%%
  %55 citations counted in INSPIRE as of 12 Feb 2017


%\cite{Gover:2014vxa}
\bibitem{Gover:2014vxa} 
  A.~R.~Gover, E.~Latini and A.~Waldron,
  %``Metric projective geometry, BGG detour complexes and partially massless gauge theories,''
  Commun.\ Math.\ Phys.\  {\bf 341}, no. 2, 667 (2016)
  doi:10.1007/s00220-015-2490-x
  [arXiv:1409.6778 [hep-th]].
  %%CITATION = doi:10.1007/s00220-015-2490-x;%%
  %4 citations counted in INSPIRE as of 12 Feb 2017


%\cite{Bernard:2014bfa}
\bibitem{Bernard:2014bfa} 
  L.~Bernard, C.~Deffayet and M.~von Strauss,
  %``Consistent massive graviton on arbitrary backgrounds,''
  Phys.\ Rev.\ D {\bf 91}, no. 10, 104013 (2015)
  doi:10.1103/PhysRevD.91.104013
  [arXiv:1410.8302 [hep-th]].
  %%CITATION = doi:10.1103/PhysRevD.91.104013;%%
  %15 citations counted in INSPIRE as of 12 Feb 2017


%\cite{Bernard:2015mkk}
\bibitem{Bernard:2015mkk} 
  L.~Bernard, C.~Deffayet and M.~von Strauss,
  %``Massive graviton on arbitrary background: derivation, syzygies, applications,''
  JCAP {\bf 1506}, 038 (2015)
  doi:10.1088/1475-7516/2015/06/038
  [arXiv:1504.04382 [hep-th]].
  %%CITATION = doi:10.1088/1475-7516/2015/06/038;%%
  %11 citations counted in INSPIRE as of 12 Feb 2017


%\cite{Bernard:2015uic}
\bibitem{Bernard:2015uic} 
  L.~Bernard, C.~Deffayet, A.~Schmidt-May and M.~von Strauss,
  %``Linear spin-2 fields in most general backgrounds,''
  Phys.\ Rev.\ D {\bf 93}, no. 8, 084020 (2016)
  doi:10.1103/PhysRevD.93.084020
  [arXiv:1512.03620 [hep-th]].
  %%CITATION = doi:10.1103/PhysRevD.93.084020;%%
  %7 citations counted in INSPIRE as of 12 Feb 2017


%\cite{Carroll:2004st}
\bibitem{Carroll:2004st} 
  S.~M.~Carroll,
  %``Spacetime and geometry: An introduction to general relativity,''
  San Francisco, USA: Addison-Wesley (2004) 513 p
  %233 citations counted in INSPIRE as of 12 Feb 2017


%\cite{Fierz:1939ix}
\bibitem{Fierz:1939ix} 
  M.~Fierz and W.~Pauli,
  %``On relativistic wave equations for particles of arbitrary spin in an electromagnetic field,''
  Proc.\ Roy.\ Soc.\ Lond.\ A {\bf 173}, 211 (1939).
  doi:10.1098/rspa.1939.0140
  %%CITATION = doi:10.1098/rspa.1939.0140;%%
  %1054 citations counted in INSPIRE as of 12 Feb 2017


%\cite{Bengtsson:1994vn}
\bibitem{Bengtsson:1994vn} 
  I.~Bengtsson,
  %``Note on massive spin-2 in curved space,''
  J.\ Math.\ Phys.\  {\bf 36}, 5805 (1995)
  doi:10.1063/1.531288
  [gr-qc/9411057].
  %%CITATION = doi:10.1063/1.531288;%%
  %28 citations counted in INSPIRE as of 12 Feb 2017


%\cite{Porrati:2000cp}
\bibitem{Porrati:2000cp} 
  M.~Porrati,
  %``No van Dam-Veltman-Zakharov discontinuity in AdS space,''
  Phys.\ Lett.\ B {\bf 498}, 92 (2001)
  doi:10.1016/S0370-2693(00)01380-0
  [hep-th/0011152].
  %%CITATION = doi:10.1016/S0370-2693(00)01380-0;%%
  %175 citations counted in INSPIRE as of 12 Feb 2017

%\cite{Nepomechie:1983yq}
\bibitem{Nepomechie:1983yq} 
  R.~I.~Nepomechie,
  %``Einstein Gravity as the Low-energy Effective Theory of Weyl Gravity,''
  Phys.\ Lett.\  {\bf 136B}, 33 (1984).
  doi:10.1016/0370-2693(84)92050-1
  %%CITATION = doi:10.1016/0370-2693(84)92050-1;%%
  %19 citations counted in INSPIRE as of 19 Apr 2017
  
  
  \cite{Buchbinder:1999ar}
\bibitem{Buchbinder:1999ar} 
  I.~L.~Buchbinder, D.~M.~Gitman, V.~A.~Krykhtin and V.~D.~Pershin,
  %``Equations of motion for massive spin-2 field coupled to gravity,''
  Nucl.\ Phys.\ B {\bf 584}, 615 (2000)
  doi:10.1016/S0550-3213(00)00389-8
  [hep-th/9910188].
  %%CITATION = doi:10.1016/S0550-3213(00)00389-8;%%
  %94 citations counted in INSPIRE as of 19 Apr 2017
  

  %\cite{Faci:2012yg}
\bibitem{Faci:2012yg} 
  S.~Faci,
  %``Constructing conformally invariant equations by using Weyl geometry,''
  Class.\ Quant.\ Grav.\  {\bf 30}, 115005 (2013)
  doi:10.1088/0264-9381/30/11/115005
  [arXiv:1212.2599 [hep-th]].
  %%CITATION = doi:10.1088/0264-9381/30/11/115005;%%
  %5 citations counted in INSPIRE as of 19 Apr 2017
  
  %\cite{Achour:2013afa}
\bibitem{Achour:2013afa} 
  J.~Ben Achour, E.~Huguet and J.~Renaud,
  %``Conformally invariant wave equation for a symmetric second rank tensor (Òspin-2Ó) in a $d$-dimensional curved background,''
  Phys.\ Rev.\ D {\bf 89}, 064041 (2014)
  doi:10.1103/PhysRevD.89.064041
  [arXiv:1311.3124 [gr-qc]].
  %%CITATION = doi:10.1103/PhysRevD.89.064041;%%
  %4 citations counted in INSPIRE as of 19 Apr 2017

%\cite{deRham:2011rn}
\bibitem{deRham:2011rn} 
  C.~de Rham, G.~Gabadadze and A.~J.~Tolley,
  %``Ghost free Massive Gravity in the St\'uckelberg language,''
  Phys.\ Lett.\ B {\bf 711}, 190 (2012)
  doi:10.1016/j.physletb.2012.03.081
  [arXiv:1107.3820 [hep-th]].
  %%CITATION = doi:10.1016/j.physletb.2012.03.081;%%
  %193 citations counted in INSPIRE as of 12 Feb 2017


%\cite{Hassan:2011ea}
\bibitem{Hassan:2011ea} 
  S.~F.~Hassan and R.~A.~Rosen,
  %``Confirmation of the Secondary Constraint and Absence of Ghost in Massive Gravity and Bimetric Gravity,''
  JHEP {\bf 1204}, 123 (2012)
  doi:10.1007/JHEP04(2012)123
  [arXiv:1111.2070 [hep-th]].
  %%CITATION = doi:10.1007/JHEP04(2012)123;%%
  %289 citations counted in INSPIRE as of 12 Feb 2017


%\cite{Hassan:2011zd}
\bibitem{Hassan:2011zd} 
  S.~F.~Hassan and R.~A.~Rosen,
  %``Bimetric Gravity from Ghost-free Massive Gravity,''
  JHEP {\bf 1202}, 126 (2012)
  doi:10.1007/JHEP02(2012)126
  [arXiv:1109.3515 [hep-th]].
  %%CITATION = doi:10.1007/JHEP02(2012)126;%%
  %389 citations counted in INSPIRE as of 12 Feb 2017


%\cite{Hassan:2011tf}
\bibitem{Hassan:2011tf} 
  S.~F.~Hassan, R.~A.~Rosen and A.~Schmidt-May,
  %``Ghost-free Massive Gravity with a General Reference Metric,''
  JHEP {\bf 1202}, 026 (2012)
  doi:10.1007/JHEP02(2012)026
  [arXiv:1109.3230 [hep-th]].
  %%CITATION = doi:10.1007/JHEP02(2012)026;%%
  %275 citations counted in INSPIRE as of 12 Feb 2017


%\cite{Hassan:2011hr}
\bibitem{Hassan:2011hr} 
  S.~F.~Hassan and R.~A.~Rosen,
  %``Resolving the Ghost Problem in non-Linear Massive Gravity,''
  Phys.\ Rev.\ Lett.\  {\bf 108}, 041101 (2012)
  doi:10.1103/PhysRevLett.108.041101
  [arXiv:1106.3344 [hep-th]].
  %%CITATION = doi:10.1103/PhysRevLett.108.041101;%%
  %464 citations counted in INSPIRE as of 12 Feb 2017


%\cite{Hassan:2011vm}
\bibitem{Hassan:2011vm} 
  S.~F.~Hassan and R.~A.~Rosen,
  %``On Non-Linear Actions for Massive Gravity,''
  JHEP {\bf 1107}, 009 (2011)
  doi:10.1007/JHEP07(2011)009
  [arXiv:1103.6055 [hep-th]].
  %%CITATION = doi:10.1007/JHEP07(2011)009;%%
  %246 citations counted in INSPIRE as of 12 Feb 2017


%\cite{Boulware:1973my}
\bibitem{Boulware:1973my} 
  D.~G.~Boulware and S.~Deser,
  %``Can gravitation have a finite range?,''
  Phys.\ Rev.\ D {\bf 6}, 3368 (1972).
  doi:10.1103/PhysRevD.6.3368
  %%CITATION = doi:10.1103/PhysRevD.6.3368;%%
  %716 citations counted in INSPIRE as of 12 Feb 2017




\bibitem{Hall}
  G.~S.~Hall, JMP,
	{\bf 32}, 181, 1991; doi: 10.1063/1.529114.
  %``Partial Masslessness and Conformal Gravity,''
  
	
	
\bibitem{Coley}
A.~A.~Coley, B.~O.~Tupper, General Relativity and Gravitation, Vol.~23, No.~10, 1991.
	
\bibitem{Levy}
H.~Levy, 
Annals of Mathematics
Second Series, Vol. 27, No. 2 (Dec., 1925), pp. 91-98.

\bibitem{Eisenhart}
Eisenhart, Transactions of the American Mathematical Society, vol. 25, p303.
	

\bibitem{Stephani:2003tm}
  H.~Stephani, D.~Kramer, M.~A.~H.~MacCallum, C.~Hoenselaers and E.~Herlt, 
  ``Exact solutions of Einstein's field equations,'' (Cambridge Monographs on Mathematical Physics).
  %%CITATION = INSPIRE-619666;%%
  %206 citations counted in INSPIRE as of 28 Feb 2017



\bibitem{xTensor}
J.-M.~Mart\'{\i}n-Garc\'{\i}a,
Comp. Phys. Commun. {\bf 179}, 597 (2008)
[arXiv:0803.0862 [cs.SC]],
$<$http://metric.iem.csic.es/Martin-Garcia/xAct/$>$.

	
\end{thebibliography}
\end{document}